\documentclass[lettersize,journal]{IEEEtran}
\usepackage{amsmath,amsfonts}
\usepackage{algorithmic}
\usepackage{algorithm}
\usepackage{array}
\usepackage[caption=false,font=normalsize,labelfont=sf,textfont=sf]{subfig}
\usepackage{textcomp}
\usepackage{stfloats}
\usepackage{url}
\usepackage{verbatim}
\usepackage{graphicx}
\usepackage{cite}
\usepackage{amssymb}
\usepackage{booktabs}

\begin{document}

\title{Radio Echo Representation Learning for Semantic Sensing in Multi-Modal ISAC Systems}

\author{
	\IEEEauthorblockN{ 
		Guanping Shang,~\IEEEmembership{Graduate Student Member,~IEEE}
        Anzheng Tang,~\IEEEmembership{Member,~IEEE}
        Kwok Kit Chan,
		and Shenghui Song,~\IEEEmembership{Senior Member,~IEEE}
	}
    \thanks{This work was supported in part by 
    the Research Grants Council of Hong Kong under the General Research Fund (Project No. 16214125).
    This manuscript is an extended version of a paper accepted by \textit{IEEE Conference on Integrated Sensing and Communication} (ISAC), Lisbon, Portugal, Nov. 2026.
    }
    \thanks{G. Shang, A. Tang and S.H. Song are with the Department of Electronic and 
    Computer Engineering, The Hong Kong University of Science and Technology, 
    Hong Kong, China. S.H. Song is also with the Division of Integrative Systems and Design, 
    The Hong Kong University of Science and Technology, Hong Kong, China
    (e-mail: gshang@connect.ust.hk; eeaztang@ust.hk; eeshsong@ust.hk).}
    \thanks{K.K. Chan is with the Capital Works Business Unit of the MTR Corporation, 
    Hong Kong, China (e-mail: kkitchan@mtr.com.hk). }

}
\maketitle

\begin{abstract}
Semantic sensing extends integrated sensing and communication (ISAC) beyond physical parameter estimation by exploiting task-oriented representations from radio echoes. 
Nevertheless, existing multimodal sensing remains largely limited to vision-assisted radio physical parameter estimation or typically processes the visual and radio modalities separately, leaving radio representations underutilized for visual task enhancement. 
Feature-level fusion enables closer cross-modal interaction, while echo interference complicates the learning of user-specific radio representations and its semantic consistency with the visual entities. 
To address these challenges, this paper proposes a multimodal semantic sensing framework for orthogonal frequency-division multiplexing (OFDM)-ISAC systems.
We propose a radio encoder to learn radio representations oriented toward visual localization from superimposed pilot echoes, together with Variational Radio-Visual Grounding (VRVG) to promote semantic consistency between the learned radio representations and visual spatial features. 
The radio representations are then formulated as semantic probes to highlight the corresponding entities in the visual scene through cross-modal fusion.
Experiments using real-world RCooper dataset show that the resulting synergistic radio-visual embeddings enable accurate radio-visual user matching and significantly improve both user-specific visual localization and physical parameter estimation performance. The results substantiate the benefits of radio representation learning and semantically consistent fusion for multimodal semantic sensing.

\end{abstract}

\begin{IEEEkeywords}
Multimodal ISAC, semantic sensing, visual detection, user matching, position and velocity estimation.
\end{IEEEkeywords}

\section{Introduction}
Integrated Sensing and Communication (ISAC) has been widely recognized as a key enabling technology for sixth-generation (6G) wireless networks.
It enables future communication networks to provide sensing services for a wide range of emerging applications, including traffic monitoring, autonomous driving, and smart transportation management \cite{review1bupt,review2hkust,review3liufan10challenges}.
In conventional ISAC systems, radio sensing based on communication waveforms, such as orthogonal frequency-division multiplexing (OFDM), primarily focuses on estimating physical parameters, including delay, angle, and Doppler shift, from which targets' position and velocity parameters can be estimated \cite{mmWave2,shang}.
In fact, radio echoes also contain implicit spatiotemporal structure of the environment.
Semantic sensing has recently emerged as a paradigm that extends radio sensing beyond physical parameter estimation to the extraction of higher-level semantic information \cite{SemS}.
By learning inference task-oriented representations from echoes, radio sensing can support higher-level inference tasks, such as action recognition and target classification \cite{widar3}.

Along another line of research, the complementary nature of radio and visual sensing has made their integration an increasingly promising research direction \cite{cameraRadarFusionReview2024,synesthesiaofmachinesChengxiang}.
Specifically, vision provides rich information about object appearance and scene context, and these high-level semantic cues have been explored to assist physical parameter estimation in radio sensing \cite{cameraRadarFusionReview2024}.
Conversely, radio echoes also encode rich environmental semantics, and their learned representations have the potential to extend the role of radio sensing beyond physical parameter estimation to supporting visual inference.
Multimodal semantic sensing that integrates radio and vision therefore offers a promising approach to more comprehensive environmental understanding \cite{synesthesiaofmachinesChengxiang}.

\IEEEpubidadjcol
\subsection{Related Works and Challenges}
Conventional and learning-based OFDM sensing methods exploit the temporal, frequency, and spatial structure of received signals to estimate physical parameters, including target position and velocity \cite{ofdmsensing2,ofdmDL3,ofdmDL4}.
Beyond parameter estimation, features extracted from OFDM observations have also been used for higher-level inference, including in-home monitoring \cite{semanticsensing1} and distributed gesture recognition \cite{semanticsensing2}.
These studies demonstrate that radio sensing can support higher-level inference tasks by encoding the implicit spatiotemporal structure of sensing channels into task-oriented semantic representations.

Extending this insight to multimodal sensing, the features embedded in radio signals are associated with the same underlying objects and scene structure observed by cameras, while providing complementary spatial and motion information \cite{cameraRadarFusionReview2024,synesthesiaofmachinesChengxiang}.
This semantic correspondence and complementarity suggest that radio sensing can actively support multimodal inference about the surrounding environment.
An established line of research, however, exploits such cooperation primarily for physical level sensing.
Specifically, visual spatial features can be mapped to geometry parameters, providing supplementary cues for radio-based localization \cite{yangjie2025slam} and prediction \cite{multimodalbeamforming2,zhangkai2025thz}.
In these applications, visual semantics mainly assist tasks in the physical parameter estimation, leaving the reciprocal contribution of radio semantic sensing to visual inference less directly explored.

To exploit cross-modal correspondence for multimodal sensing, Guo \textit{et al.} propose a semantic-aware monostatic ISAC framework for radio-vision integration \cite{guojiaxiang2025semantic}, while some studies connect radio beam or echoes with visually detected objects to infer user positions and establish user-object matching\cite{radarToUserAssociation, userIdentification}. 
These efforts extend radio sensing from estimating physical parameters toward interpreting visual entities in the environment. Nevertheless, in decoupled radio-vision pipelines, radio observations are first processed to estimated positions or other intermediate outputs, which are subsequently associated with independently obtained visual detections \cite{userIdentification,yangjie2025slam}.
Cross-modal interaction therefore takes place at the output level, leaving the underlying radio and visual representations largely separate.
Although these approaches establish object correspondence, independently processing radio signals limits the extent to which task-oriented radio representations can guide the learning and refinement of visual representations. As a result, the complementary information embedded in radio observations remains underutilized for visual inference.
Moreover, such post-alignment paradigms may introduce ambiguity,
redundancy, and cardinality mismatches in multi-object
scenarios, thereby undermining semantic correspondence between radio and visual modalities.

Feature-level fusion offers a direct way to exploit the complementarity between radio and visual modalities.
For example, SIMAC \cite{SIMAC} uses feature-level cross-attention between single-target echoes and visual semantic features to identify the visual object and estimate physical parameters.
However, for multiple user visual localization based on radio echoes, effective fusion also requires radio representations that distinguish the specific user within superimposed reflections and align with the corresponding visual entities.
This calls for a framework that learns multimodal task-oriented radio representations that are semantically consistent with their corresponding visual entities, enabling radio representations to refine visual representations and improve multimodal localization performance.

\subsection{Contributions}
To address these challenges, we propose a multimodal semantic sensing framework in OFDM-ISAC systems that extends radio semantic sensing to multimodal settings. By learning radio representations oriented toward visual localization, the framework promotes semantic consistency between radio and visual representations. Such semantic consistency facilitates both explicit user-specific matching between each ISAC user and its corresponding visual object and implicit feature-level multimodal fusion, thereby improving visual localization and physical parameter estimation under multi-object echo interference.
Table~\ref{tab:comparison} summarizes its coverage of tasks relative to representative related works.
The main contributions are summarized as follows:
\begin{itemize}
\item \textbf{Multimodal ISAC for Semantic Sensing:}
We formulate a multimodal semantic sensing paradigm for OFDM-ISAC systems, where superimposed OFDM pilot echoes are encoded into user-specific radio representations oriented toward visual localization. By using these radio representations as semantic probes to interact with visual tokens through a cross-modal attention, the proposed paradigm associates each ISAC user with its corresponding visual entity, simultaneously supporting visual localization and parameter estimation. 

\item \textbf{Radio Encoding and Semantic Consistency:}
We develop a radio echo encoder with gated compression and multi-user interaction to extract user-specific radio features from superimposed echoes.
Variational Radio-Visual Grounding (VRVG) is proposed to guide the radio representation aligned with the spatial relationships of visual entities, thereby promoting the semantic consistency between radio and vision.
Such grounded radio representations provide user-specific guidance to the fusion network, refining the visual features and improving localization performance.

\item \textbf{Radio--Visual Inference Gains:}
Experiments based on the real-world RCooper dataset \cite{RCooper} demonstrate reliable radio--vision user matching under clutter interference from other users and non-user reflectors, together with improved user-specific visual localization and physical parameter estimation performance. 
Comparisons with existing baselines and component ablations validate the advantages of radio representation learning and semantically consistent multimodal fusion. 
\end{itemize}

\begin{table}[t]
\caption{Coverage of Sensing Topics in Representative Works}
\label{tab:comparison}
\centering
\small
\setlength{\tabcolsep}{3pt}
\renewcommand{\arraystretch}{1.1}
\begin{tabular}{@{}l*{7}{c}@{}}
\toprule
\textbf{Topic} & \textbf{Ours} & \cite{SemS} & \cite{ofdmDL3} & \cite{yangjie2025slam} & \cite{userIdentification} & \cite{SIMAC} & \cite{gaofeifei}\\
\midrule
OFDM ISAC & $\checkmark$ &    & $\checkmark$ &    &    &    &   \\
Multimodal sensing &     &    &    & $\checkmark$ & $\checkmark$ & $\checkmark$ & $\checkmark$\\
Semantic sensing & $\checkmark$ & $\checkmark$ &    &    & $\checkmark$ & $\checkmark$ &    \\
Radio-vision matching & $\checkmark$ &    &    & $\checkmark$ & $\checkmark$ &    & $\checkmark$\\
Parameter estimation & $\checkmark$ & $\checkmark$ & $\checkmark$ & $\checkmark$ &    & $\checkmark$ & $\checkmark$\\
\bottomrule
\end{tabular}
\vspace{-5pt}
\end{table}

\subsection{Paper Structure and Notations}
\subsubsection*{Paper Structure}
The remainder of this paper is organized as follows.
Section II presents the OFDM-ISAC system and formulates the user-specific semantic sensing problem.
Section III describes radio representation learning and radio--visual grounding.
Section IV presents multimodal semantic fusion, the joint objective, and the training strategy.
Section V reports the experimental results, and Section VI concludes the paper.
\subsubsection*{Notations}
Throughout this paper, scalars are denoted by italic letters (e.g., $a$),
vectors by bold lowercase letters (e.g., $\mathbf{a}$),
and tensors by bold uppercase letters (e.g., $\mathbf{A}$).
$\mathbb{R}$ and $\mathbb{C}$ denote the sets of real and complex numbers, respectively.
$(\cdot)^{T}$ and $(\cdot)^{H}$ represent the transpose and Hermitian transpose, respectively.
$\|\cdot\|_p$ denotes the $\ell_p$ norm of a vector and
$\mathbb{E}(\cdot)$ represents statistical expectation. $\circledast$ denotes 3D convolution, $\oslash$ denotes element-wise division, and $\odot$ denotes element-wise multiplication.

\section{System Model and Semantic Sensing Problem}
\label{sec:system_model}

\begin{figure}[t]
\centering
\includegraphics[width=3.2in]{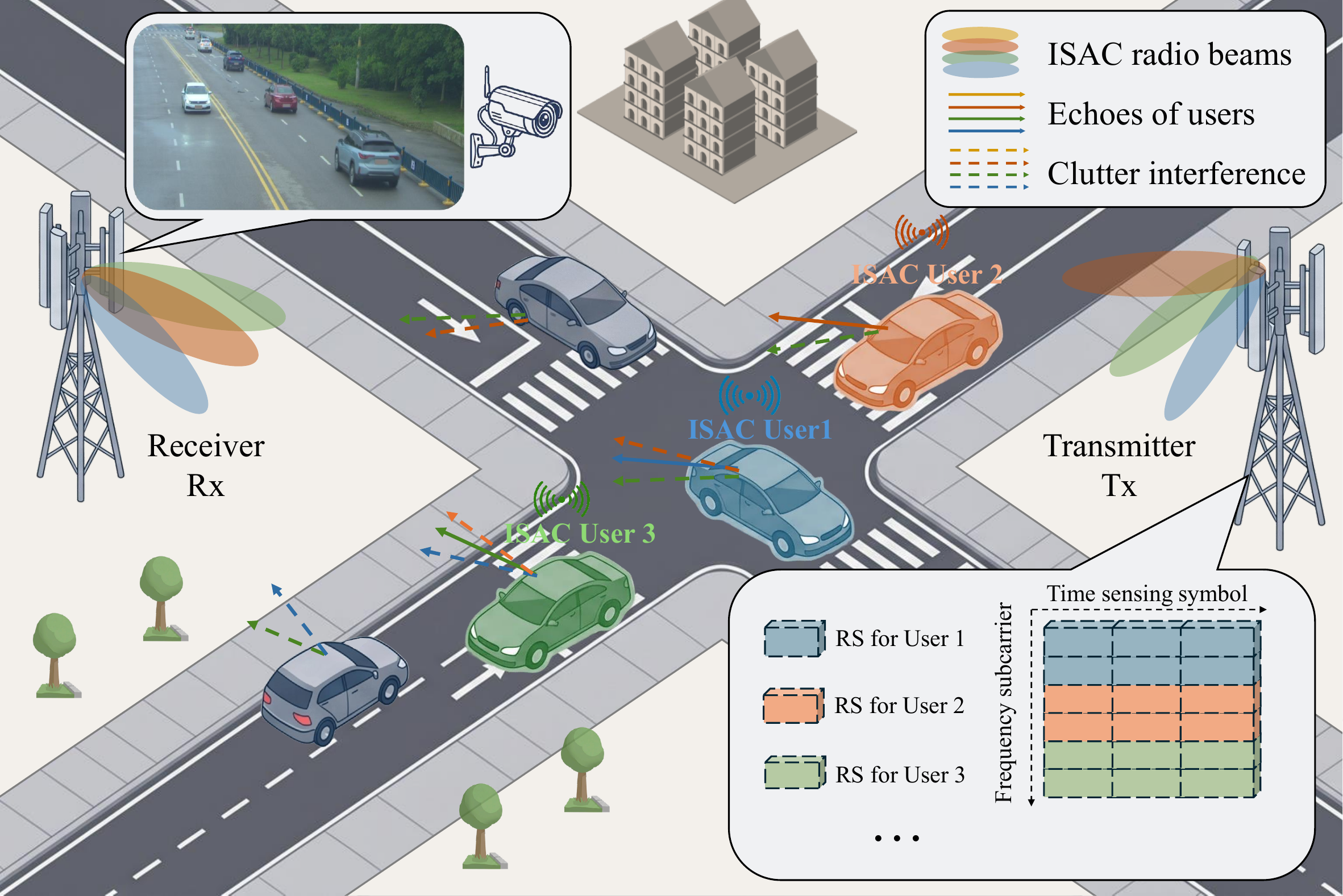}
\caption{Bistatic OFDM-ISAC system with user-specific pilot resources. 
The received pilot signal contains the intended-user echo together with superimposed echoes from other reflectors.}
\label{scenario}
\end{figure}

Consider the bistatic OFDM-ISAC system in Fig.~\ref{scenario}, comprising a transmitter Tx and a camera-equipped receiver Rx. 
Tx provides downlink communication services to the ISAC users and reuses their orthogonal pilots for sensing, such that the served communication users also act as sensing targets.
Rx jointly processes the corresponding pilot echoes with visual images, where the echoes combine the intended-user return with reflections from surrounding users and non-user objects. 
The objective of the system is to learn user-specific radio representations from the superimposed observations for multimodal inference, supporting radio--visual user matching and visual localization while jointly estimating user position and velocity.

\subsection{Signal Model}
\label{signal model_chapter}
Let $\mathcal U=\{1,\ldots,U\}$ denote the set of $U$ served ISAC users, with user $i\in\mathcal U$ assigned an orthogonal pilot frequency band $B_i$.
Each band contains $N_s$ pilot subcarriers with spacing $\Delta f$, observed over a coherent interval spanning over $T_s$ OFDM symbols.
Let $k=0,\ldots,N_s-1$ and $\ell=0,\ldots,T_s-1$ denote the local subcarrier and symbol indices, respectively, and let $T_{\rm sym}$ denote the symbol interval.
With $N_t$ transmit antennas, Tx applies a unit-norm beam $\mathbf w_i\in\mathbb C^{N_t\times1}$ to the known nonzero pilot symbol $S_i[k,\ell]\in\mathbb C$.
The antenna-domain slices of the transmitted tensor $\mathbf X_i\in\mathbb C^{N_t\times N_s\times T_s}$ are given by  
\begin{equation}
\label{xeqws}
\mathbf X_i [k,\ell]=\mathbf w_i S_i[k,\ell].
\end{equation}

\begin{figure}[t]
\centering
\includegraphics[width=2.7in]{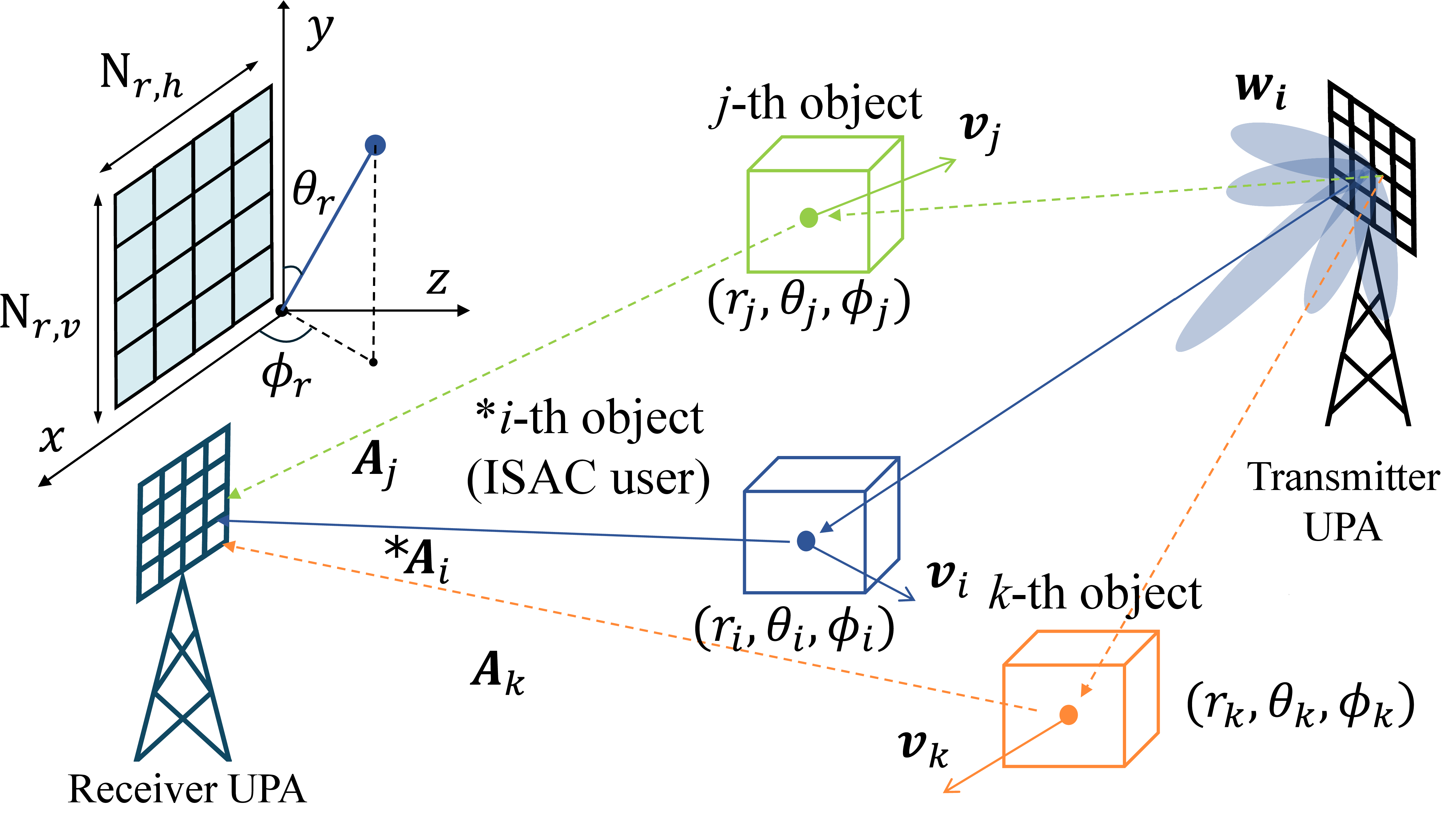}
\caption{Reflection model for the UPA array. The beam toward the $i$-th served ISAC user also illuminates surrounding objects.}
\label{reflectmodel}
\end{figure}

To characterize the spatial structure of the sensing echoes, we first describe the array responses.
As shown in Fig.~\ref{reflectmodel}, Tx and Rx employ half-wavelength spaced uniform planar arrays (UPAs). Tx is equipped with $N_t = N_{t,v} \times N_{t,h}$ antennas, 
where $N_{t ,v}$ and $N_{t,h}$ denote the number of antennas in the vertical and horizontal dimensions, respectively.
The vertical and horizontal array steering vectors are respectively given by 
\begin{equation}
\mathbf b_{t,v}(\theta_t)=
\frac{1}{\sqrt{N_{t,v}}}
[1,\ldots,e^{-\mathrm j(N_{t,v}-1)\pi\cos\theta_t}]^T,
\end{equation}
\begin{equation}
\mathbf b_{t,h}(\phi_t,\theta_t)=
\frac{1}{\sqrt{N_{t,h}}}
[1,\ldots,e^{-\mathrm j(N_{t,h}-1)\pi\sin\theta_t\cos\phi_t}]^T,
\end{equation}
where $\theta_{t}$ and $\phi_{t}$ denote the elevation and azimuth angles of the user 
with respect to the base station, respectively\cite{UPA}. 
The transmit steering vector is expressed as the Kronecker product of the vertical and horizontal steering vectors 
\begin{equation}
\label{UPAsteeringvector}
\mathbf a_t(\phi_t,\theta_t)=
\mathbf b_{t,v}(\theta_t)\otimes\mathbf b_{t,h}(\phi_t,\theta_t) \in\mathbb C^{N_t\times1}, 
\end{equation}
and the receiving steering vector $\mathbf{a}_r(\phi_r, \theta_r) \in \mathbb{C}^{N_r \times 1}$ is defined similarly.

Although $\mathbf w_i$ directs pilot energy toward user $i$, its finite main-lobe width and side lobes also illuminate other reflectors. 
The illuminated scene comprises $O$ effective reflectors, where $\mathcal R=\{U+1,\ldots,O\}$ denotes the set of non-user reflectors contributing environmental echoes.
The spatial sensing channel via reflector $j$ is given by 
\begin{equation}
\label{sensingchannelAeq}
\mathbf A_j=\beta_j
\mathbf a_r(\phi_{r,j},\theta_{r,j})
\mathbf a_t^H(\phi_{t,j},\theta_{t,j}) \in \mathbb{C}^{N_{r} \times N_{t}} ,
\end{equation}
where $\mathbf{\beta}_j$ denotes the effective complex scattering coefficient accounting for propagation loss and radar cross section modeled as complex Gaussian\cite{sensingchannelA}. The band index is suppressed in $\mathbf A_j$ as its constant frequency-offset phase is absorbed into $\beta_j$. 

Beyond spatial structure, the echoes carry delay and Doppler cues determined by reflector position and velocity.
For reflector $j$, the bistatic delay and Doppler shift are given by 
\begin{equation}
\label{eq:bistatic_delay}
\tau_j=(\|\mathbf p_j-\mathbf p_t\|_2+
\|\mathbf p_j-\mathbf p_r\|_2)/c,
\end{equation}
\begin{equation}
\label{eq:bistatic_doppler}
\nu_j=\frac{\mathbf v_j^T}{\lambda}
\left(\frac{\mathbf p_t-\mathbf p_j}{\|\mathbf p_t-\mathbf p_j\|_2}
+\frac{\mathbf p_r-\mathbf p_j}{\|\mathbf p_r-\mathbf p_j\|_2}\right),
\end{equation}
where $c$ and $\lambda$ denote the speed of light and the carrier wavelength, respectively, $\mathbf v_j$ represents the velocity of reflector $j$, and $\mathbf p_t$, $\mathbf p_r$, and $\mathbf p_j$ specify the three-dimensional positions of Tx, Rx, and reflector $j$, respectively.

After timing and carrier synchronization, Rx separates the user-specific frequency bands through demodulation. The received pilot echo  
$\mathbf Y_i\in\mathbb C^{N_r\times N_s\times T_s}$ corresponding to the
$i$-th user is superimposed with echoes from other users and non-user reflectors. Specifically, define the echo contributed by the $j$-th reflector under the
$i$-th user-specific pilot as
\begin{equation}
\mathbf E_{i,j}[k,\ell]
\triangleq
\mathbf A_j\mathbf X_i [k,\ell]
e^{-\mathrm j2\pi k\Delta f\tau_j}
e^{\mathrm j2\pi\ell T_{\rm sym}\nu_j}.
\end{equation}
The received echo of the $i$-th ISAC user is given by 
\begin{equation}
\label{eq:multi_scatterer_echo}
\begin{aligned}
\mathbf Y_i[k,\ell]
=&
\underbrace{\mathbf E_{i,i}[k,\ell]}_{\text{intended-user echo}}
+
\underbrace{\sum_{j\in\mathcal U\setminus\{i\}}
\mathbf E_{i,j}[k,\ell]}_{\text{other-user echoes}}
\\
&+
\underbrace{\sum_{j\in\mathcal R}
\mathbf E_{i,j}[k,\ell]}_{\text{non-user-reflector echoes}}
+
\underbrace{\mathbf z_i[k,\ell]}_{\text{receiver noise}},
\end{aligned}
\end{equation}
where $\mathbf z_i[k,\ell]\sim
\mathcal{CN}(\mathbf 0,\sigma_i^2\mathbf I_{N_r})$ represents the additive white Gaussian noise at the receiver\cite{signalphase}.

\subsection{Radio-Visual Semantic Sensing Framework}
The other-user and non-user-reflector echoes in \eqref{eq:multi_scatterer_echo} arise from the same frequency band and interfere with the intended-user echo. 
Although these clutter echoes obscure the intended user's individual response, the joint angular, delay, and Doppler structure of $\mathbf Y_i$ also contains its motion and spatial relationships with surrounding reflectors.
These radio features encode the user-specific reflection with spatial context, which provide complementary semantic cues for radio-visual user matching and visual refinement.
We thus exploit this property by extending unimodal radio semantic sensing to multimodal inference through learning user-specific representations oriented toward visual localization.

\begin{figure}[t]
\centering
\includegraphics[width=3.4in]{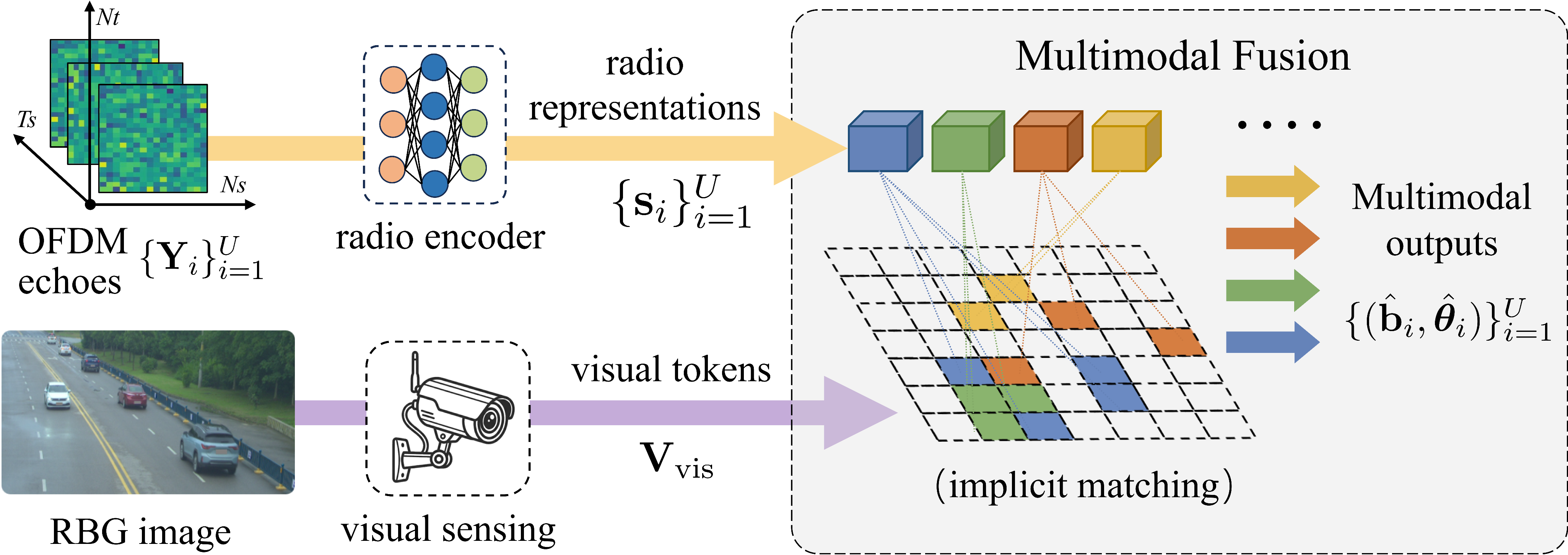}
\caption{Workflow of the proposed radio--visual semantic sensing system. 
The radio representations learned from pilot echoes are used for user-specific visual localization and parameter estimation, where the framework implicitly matches each served ISAC user with its visual entity.} 
\label{workflow}
\end{figure}

Specifically, the proposed framework comprises a radio echo encoder, a visual encoder, and a multimodal semantic fusion network, as illustrated in Fig.~\ref{workflow}.
Let $\mathbf V\in\mathbb R^{H\times W\times 3}$ denote the RGB image temporally aligned with the radio observations, where $H$ and $W$ denote the image height and width, respectively.
The radio echo encoder $\psi_{\mathrm{radio}}$ jointly processes the received echo tensors $\{\mathbf Y_i\}_{i=1}^{U}$ to extract user-specific radio representations that capture task-oriented spatial characteristics.
Meanwhile, the visual encoder $\psi_{\mathrm{vis}}$ extracts appearance features from the synchronized image $\mathbf V$.
The two encoding processes are expressed as
\begin{equation}
\label{eq:radio_semantic_encoding}
\{\mathbf s_i\}_{i=1}^{U}
=
\psi_{\mathrm{radio}}
\left(
\{\mathbf Y_i\}_{i=1}^{U};
\boldsymbol\xi_{\mathrm{radio}}
\right),
\end{equation}
\begin{equation}
\label{eq:visual_semantic_encoding}
\mathbf V_{\mathrm{vis}}
=
\psi_{\mathrm{vis}}
\left(
\mathbf V;
\boldsymbol\xi_{\mathrm{vis}}
\right),
\end{equation}
where $\mathbf s_i$ denotes the learned radio semantic representation of the $i$-th user, $\mathbf V_{\mathrm{vis}}$ denotes the visual feature representation, and $\boldsymbol\xi_{\mathrm{radio}}$ and $\boldsymbol\xi_{\mathrm{vis}}$ denote the learnable parameters of the radio and visual encoders, respectively.

The multimodal fusion network
$\psi_{\mathrm{fus}}$
utilizes the learned radio representations as semantic probes to retrieve and refine
corresponding visual features, thereby integrating complementary information
across the two modalities for multimodal inference.
The primary visual task is user-specific semantic sensing, which identifies
the visual entity corresponding to its ISAC user and predicts its
visual boundary in the image.
In parallel, the framework estimates the physical parameters of each user, including
its position and velocity.
The joint multimodal sensing outputs are expressed as
\begin{equation}
\label{eq:semantic_sensing_outputs}
\begin{aligned}
\{(\hat{\mathbf b}_i,\hat{\boldsymbol\theta}_i)\}_{i=1}^{U}
=\psi_{\mathrm{fus}}\bigl(
\{\mathbf s_i\}_{i=1}^{U},
\mathbf V_{\mathrm{vis}};\boldsymbol\xi_{\mathrm{fus}}\bigr),
\end{aligned}
\end{equation}
where $\boldsymbol\xi_{\mathrm{fus}}$ denotes the learnable parameters of the
fusion network,
$\hat{\mathbf b}_i\in\mathbb R^4$ represents the predicted bounding box of the $i$-th user visual entity\cite{zuizaobox}, and
$\hat{\boldsymbol\theta}_i=
[\hat{\mathbf p}_i^T,\hat{\mathbf v}_i^T]^T\in\mathbb R^6$ denotes the estimated position and velocity. 
Since each echo is indexed by $i$-th user, the visual localization simultaneously establishes the implicit radio--visual matching for each user.
The radio and visual encoders, as well as the multimodal semantic fusion mechanism,
are detailed in Sections~III and IV, respectively.

\section{Multimodal Semantic Feature Extraction}
\label{sec:feature_extraction}

In this section, we present the radio representation learning for user-specific multimodal inference. 
The radio branch combines gated attention compression with Variational Radio-Visual Grounding: compression produces compact tokens from superimposed OFDM observations, while VRVG uses visual spatial relationship to shape the radio representations before fusion. 
A standard visual encoder provides complementary appearance features for the subsequent radio-conditioned localization.

\subsection{Gated Attention Compression}
\begin{figure*}[t]
\centering
\includegraphics[width=0.85\textwidth]{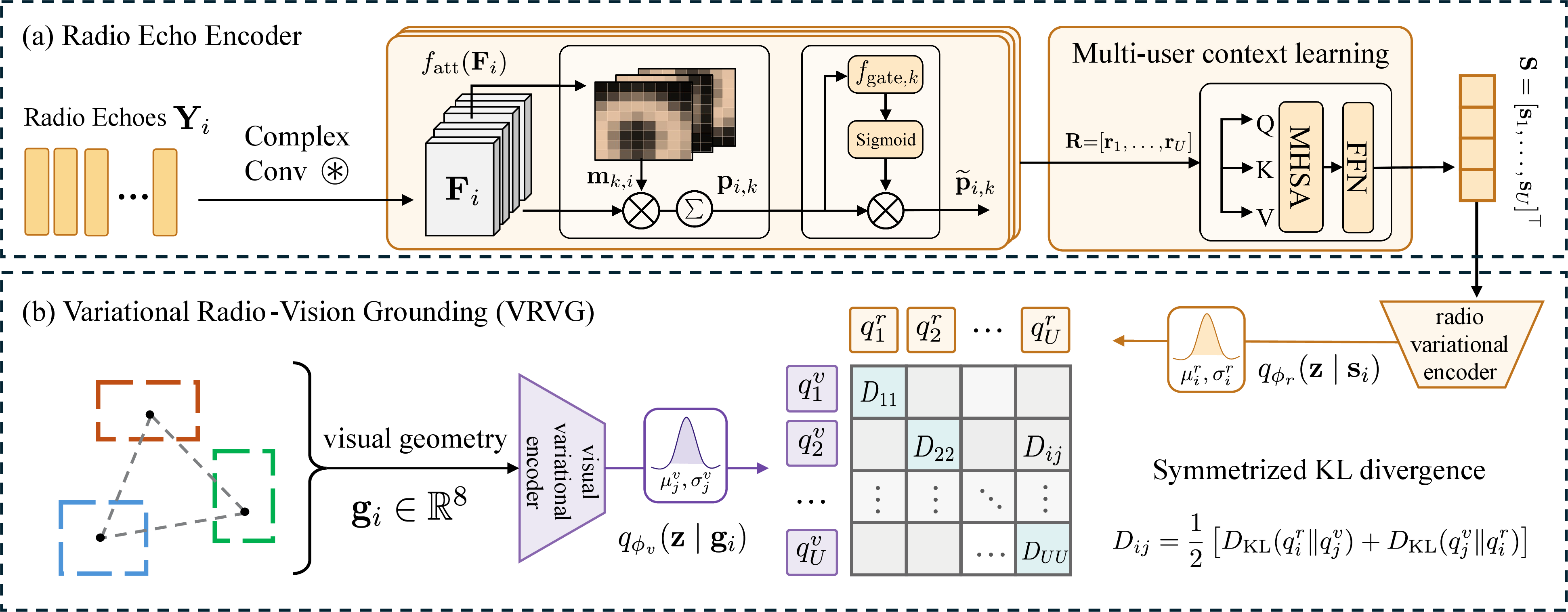}
\caption{Radio representation learning with gated attention compression and VRVG modules. 
(a) Gated attention compression extracts compact radio tokens from the pilot echoes, followed by multi-user context learning to incorporate contextual relationships among served users. 
(b) VRVG guides radio representation learning towards visual localization task through same-user distributional alignment and cross-user separation.}
\label{signalencoder}
\end{figure*}

The received radio echoes are high-dimensional across the antenna, subcarrier, and symbol axes, although its spatiotemporal structure is governed by sparse propagation components with noise.
Therefore, a compact encoding is desired to extract the discriminative sensing semantics that
preserve user-specific characteristics while suppressing interference.
To this end, we propose a convolution gated attention module as illustrated in the upper part of Fig.~\ref{signalencoder}, which effectively captures the informative features from the high-dimensional radio echoes.

Specifically, the received echoes $\mathbf Y_i$ are first normalized
by the known pilots $\mathbf X_i$ through element-wise least-squares (LS) estimation for channel observation. The resulting complex tensor is then processed by a complex convolution for subsequent feature extraction in real-valued network \cite{complexcnn}. This process is expressed as
\begin{equation}
\label{eq:complex_echo_conv}
\begin{bmatrix}\mathbf F_{\Re,i}\\ \mathbf F_{\Im,i}\end{bmatrix}
=
\begin{bmatrix}
\mathbf W_{\Re}&-\mathbf W_{\Im}\\
\mathbf W_{\Im}&\mathbf W_{\Re}
\end{bmatrix}
\circledast
\begin{bmatrix}
\Re(\mathbf Y_i\oslash\mathbf X_i)\\
\Im(\mathbf Y_i\oslash\mathbf X_i)
\end{bmatrix},
\end{equation}
where $\mathbf W_{\Re}$ and $\mathbf W_{\Im}$ denote the learnable real and
imaginary convolution kernels with C/2 channel dimensions, respectively. The outputs are concatenated into $\mathbf F_i\in\mathbb R^{C\times N_r\times N_s\times T_s}$.

Inspired by the input-adaptive tokenization\cite{ryoo2021tokenlearner}, we aggregate informative features into compact tokens using learned attention maps. Specifically, pointwise 3D convolutions layers $f_{\mathrm{att}} $ with K output channels are adopted to generate the K attention maps from $\mathbf F_i$. The output $\mathbf M_i =f_{\mathrm{att}}(\mathbf F_i) \in\mathbb R^{K\times N_r\times N_s\times T_s}$  is then normalized to $\mathbf m_{k,i}$ with
a temperature-softmax to enhance its selectivity 
\begin{equation}
\label{eq:attn_gen}
\mathbf m_{k,i}
=\operatorname{Softmax}\!\left(\mathbf M_i(k,:,:,:)/\tau
\right)
\in\mathbb R^{N_r\times N_s\times T_s},
\end{equation}
where $\tau>0$ is the temperature parameter to control the sharpness
of the attention distribution. 
The pooled token $\mathbf p_{i,k}\in\mathbb R^{1\times C}$ corresponding to the $k$-th attention map is then computed by applying the attention weights to all $C$ channels 
\begin{equation}
\label{eq:token_pool}
\mathbf p_{i,k}
=\sum_{r=1}^{N_r}\sum_{n=1}^{N_s}\sum_{l=1}^{T_s}
\mathbf m_{k,i}(r,n,l)\,
\mathbf F_i(:,r,n,l)^T.
\end{equation}
Thus, each attention weighting produces one channel-wise summary over the three observation dimensions.
The $K$ learned attention weightings provide distinct pooling patterns, allowing the compressed representation to emphasize informative features of reflection components.

We then apply channel-wise gating to suppress redundant or interference-dominated components in the pooled features. The gated feature $\widetilde{\mathbf p}_{i,k}$ is given by 
\begin{equation}
\label{eq:token_gate}
\widetilde{\mathbf p}_{i,k}
=\mathbf p_{i,k}\odot
\operatorname{Sigmoid}\bigl(f_{\mathrm{gate},k}(\mathbf p_{i,k})\bigr)\in\mathbb R^{1\times C},
\end{equation}
where $f_{\mathrm{gate},k}: \mathbb R^{1\times C}\rightarrow\mathbb R^{1\times C}$ denote the multilayer perceptron (MLP) that generates gating scores for the $k$-th pooling slot.
The sigmoid gates provide continuous, input-dependent attenuation, allowing each pooling slot to preserve distinct combinations of features. The subsequent task and grounding objectives jointly learn which components should survive compression, linking this selection mechanism to the visual spatial semantics developed next.

Finally, the gated features are concatenated and jointly projected into a compact user-specific radio feature $\mathbf r_i$ with 
\begin{equation}
\label{eq:token_compression}
\mathbf r_i 
=f_{\mathrm{cmp}}\left(
\operatorname{Concat}_{k=1}^{K}\widetilde{\mathbf p}_{i,k}\right)
\in\mathbb R^{1\times D},
\end{equation}
where $f_{\mathrm{cmp}}:\mathbb R^{1\times KC}\rightarrow\mathbb R^{1\times D}$ is an linear projection.

\subsection{Variational Radio-Visual Grounding}
\label{sbsc:sensingencoder}
\label{sec:vrvg}

Building on the compressed radio tokens, we further seek to learn task-oriented radio representations that provide complementary spatial information for visual inference. Grounded Language--Image Pre-training
(GLIP) reformulates object detection as phrase grounding through word-vision alignment, thereby learning text-aware visual representations \cite{Li2022GLIP}. Inspired by this textual grounding, the radio representation can similarly be aligned as a user-specific semantic cue that is grounded to its
corresponding visual entity.

Specifically, despite arising from different observation domains, radio echoes and visual objects describe the same physical scene and therefore share latent spatial structure. The joint angular, delay, and
Doppler patterns of the received echoes characterize not only the intended user's physical parameters but also its spatial and motion relationships with surrounding reflectors. 
The underlying scene geometry thus induces learnable cross-modal correspondences
between radio and the position and spatial extent of visual entities. Such spatial associations between radar and visual observations have been exploited for object detection and matching in radar sensing \cite{Long2021RadarCamera,Hwang2022CramNet}. 
In this subsection, we propose the VRVG to learn the latent spatial structure of radio representations and ground it in the corresponding visual semantics for radio--vision semantic sensing.

\subsubsection{Multi-user context learning}

As discussed above, radio representations for visual inference depend on both the intended-user reflection and its spatial relationships with surrounding reflectors. Since the user-specific pilots probe the same scene under different beam directions, their echo tokens provide complementary observations of shared reflectors and inter-user geometry. We therefore introduce a multi-user context learning module to aggregate such complementary information for subsequent radio--vision grounding.

Stacking the compressed radio features gives
$\mathbf R=[\mathbf r_1,\ldots,\mathbf r_U]^\top\in\mathbb R^{U\times D}$,
where each row corresponds to one user-specific radio token.
We employ multi-head self-attention (MHSA) to aggregate complementary
information across these tokens \cite{vaswani2017attention}. For the $h$-th attention head, 
\begin{equation}
\label{eq:radio_mhsa}
\begin{aligned}
&\mathbf H^r_h=\operatorname{Softmax}
\left(\mathbf Q^r_h {\mathbf K^r_h}^T/\sqrt{d^r}\right)\mathbf V^r_h \in\mathbb R^{U\times d^r} ,\\
&\operatorname{MHSA}(\mathbf R)
=\operatorname{Concat}_{h=1}^{N^r}(\mathbf H^r_h)\mathbf W^r\in\mathbb R^{U\times D},
\end{aligned}
\end{equation}
where $\mathbf Q^r_h=\mathbf R\mathbf W^r_{Q,h}$, $\mathbf K^r_h=\mathbf R\mathbf W^r_{K,h}$, and $\mathbf V^r_h=\mathbf R\mathbf W^r_{V,h}$ denote the query, key, and value matrices of the $h$-th attention head, respectively. 
$N^r$ is the number of attention heads, $d^r=D/N^r$ is the dimension of each head, and $\mathbf W^r\in\mathbb R^{D\times D}$ is a linear projection for concatenated outputs.
The following
residual connections and a feedforward network (FFN) with layer normalization (LN) \cite{vaswani2017attention} are applied to each user-specific token as
\begin{equation}\begin{aligned}
\widetilde{\mathbf R}
=&
\operatorname{LN}\!\left(
\mathbf R+\operatorname{MHSA}(\mathbf R)
\right)\in\mathbb R^{U\times D}.
\\
\mathbf S
=&
\operatorname{LN}\!\left(
\widetilde{\mathbf R}
+
\operatorname{FFN}(\widetilde{\mathbf R})
\right)
\in\mathbb R^{U\times D},
\end{aligned}
\end{equation}
where
$\mathbf S=[\mathbf s_1,\ldots,\mathbf s_U]^\top$
and $\mathbf s_i\in\mathbb R^{D}$ denotes the context-learned radio representations corresponding to the $i$-th user. These context-learned radio representations are then used for subsequent radio--vision grounding and multimodal fusion.

\subsubsection{Visual geometry}

To ground the radio representation toward visual boundary, the visual
modality should provide a representation that depends not only on the physical parameters, but also
on its spatial relations with other reflectors in the scene. These relations
provide a natural counterpart to the inter-user context aggregated
in the radio representations. We therefore construct a visual spatial
descriptor that characterizes both absolute object geometry and relative
scene structure.

Specifically, let $\mathbf b_i=[c_{x,i},c_{y,i},w_i,h_i]$ denote the normalized visual box of user $i$. The pixel center $\mathbf c_i=[c_{x,i},c_{y,i}]$ with mean $\bar{\mathbf c}$ and size $\mathbf d_i=[w_i,h_i]$ with mean $\bar{\mathbf d}$ for $U$ served users in the current scene.
The visual spatial descriptor is then constructed as 
\begin{equation}
\label{eq:vrvg_geometry}
\mathbf g_i=
\left[
\mathbf b_i,\;
\mathbf c_i-\bar{\mathbf c},\;\mathbf d_i\oslash\bar{\mathbf d}
\right]\in\mathbb R^8. 
\end{equation}
The absolute boundary $\mathbf b_i$ anchors the visual entity in the image,
whereas the relative center and scale terms characterize its spatial
features with respect to the current user set. Accordingly,
$\mathbf g_i$ provides the visual-side spatial representation to ground the corresponding radio representation. 
Only the $U$ associated visual entities are available to construct this grounding signal, while non-user reflectors remain present in the radio observations through \eqref{eq:multi_scatterer_echo}.

\subsubsection{Variational Radio-Visual Grounding}
The context-learned radio representations $\mathbf s_i$ and visual spatial descriptor $\mathbf g_i$ provide modality-specific representations of the spatial semantics associated with the same user. We therefore seek to strengthen the consistency between the matched radio--vision pair while suppressing its correlation with other users. For a fixed user count $U$, this objective can be expressed in terms of mutual information as
\begin{equation}
\label{eq:vrvg_mi_goal}
\begin{aligned}
\mathcal J_{\mathrm{MI}}
=\frac1U\sum_{i=1}^{U}I(\mathbf s_i;\mathbf g_i)
-\frac{1}{U(U-1)}
\sum_{i=1}^{U}\sum_{j\ne i}
I(\mathbf s_i;\mathbf g_j).
\end{aligned}
\end{equation}
Maximizing $\mathcal J_{\mathrm{MI}}$ promotes stronger semantic consistency between radio and vision modalities of the same user, thereby encouraging user-discriminative alignment.

Directly optimizing \eqref{eq:vrvg_mi_goal} requires unknown cross-modal joint and conditional densities. 
Vision--language alignment methods typically learn shared semantics through deterministic embeddings and an InfoNCE-type contrastive objective\cite{radford2021clip}, which provides a natural starting point for cross-modal correspondence learning.
For radio--vision sensing, however, a user-specific representation also varies with the surrounding reflectors, beam-dependent illumination, and sensing noise.
Pointwise similarity may bias alignment toward low-dimensional physical parameters, such as position, rather than preserving the richer variation of the underlying spatial semantics. 
We therefore seek a distributional representation that can align user-specific spatial structure across these varying observations, while retaining the relative geometry needed for visual localization.

Specifically, we model the radio and visual representations as user-specific conditional Gaussian embeddings. For the $i$-th user, let $q_{\phi_r}(\mathbf z\mid\mathbf s_i)$ and $q_{\phi_v}(\mathbf z\mid\mathbf g_i)$ denote the conditional latent distributions produced by the radio and visual embedding networks with parameters $\phi_r$ and $\phi_v$, respectively. Following the diagonal-Gaussian parameterization \cite{kingma2014vae,vilnis2015gaussian}, we have
\begin{equation}
\label{eq:vrvg_gaussians}
\begin{aligned}
q_i^r(\mathbf z)
&=q_{\phi_r}(\mathbf z\mid\mathbf s_i)
=\mathcal N\!\left(
\boldsymbol\mu_i^r,
\operatorname{diag}[(\boldsymbol\sigma_i^r)^2]
\right),\\
q_i^v(\mathbf z)
&=q_{\phi_v}(\mathbf z\mid\mathbf g_i)
=\mathcal N\!\left(
\boldsymbol\mu_i^v,
\operatorname{diag}[(\boldsymbol\sigma_i^v)^2]
\right).
\end{aligned}
\end{equation}
where $\mathbf z$ denotes the shared latent variable,
$q_i^r$ and $q_i^v$ are the resulting radio and visual Gaussian
embedding distributions, and
$\boldsymbol\mu_i^{r(v)}$ and $\boldsymbol\sigma_i^{r(v)}$
denote their latent mean and standard deviation, respectively.

To connect the latent distributions with the mutual-information objective
in \eqref{eq:vrvg_mi_goal}, consider the conditional distribution $p_v(\mathbf z_i\mid\mathbf s_i)$, 
introduced by \(p(\mathbf s_i,\mathbf g_i)q_{\phi_v}(\mathbf z_i|\mathbf g_i)\), where $\mathbf z_i$ is generated from $\mathbf g_i$ through the visual embedding $q_{\phi_v}$. 
Since this induced $p_v(\mathbf z_i\mid\mathbf s_i)$ is generally intractable, we approximate it
using the radio embedding distribution 
$q_{\phi_r}(\mathbf z_i\mid\mathbf s_i)$. 
Applying the Barber--Agakov variational lower bound
\cite{barber2003mi,poole2019mi} gives 
\begin{equation}
\begin{aligned}
I(\mathbf s_i;\mathbf g_i) \ge
I(\mathbf s_i;\mathbf z_i)
&=
h(\mathbf z_i)
+
\mathbb E
\left[
\log p_v(\mathbf z_i\mid\mathbf s_i)
\right]\\
&\ge
h(\mathbf z_i)
+
\mathbb E
\log q_{\phi_r}(\mathbf z_i\mid\mathbf s_i),
\end{aligned}
\label{eq:vrvg_mi_sz}
\end{equation}
Since $\mathbf z_i$ is sampled from
$q_{\phi_v}(\mathbf z_i\mid\mathbf g_i)$, the variational term can be
rewritten as the divergence between the radio and visual Gaussian embeddings with
\begin{equation}
\begin{aligned}
\mathbb E\log q_{\phi_r}(\mathbf z_i\mid\mathbf s_i)=
\mathbb E\log q_{\phi_v}(\mathbf z_i\mid\mathbf g_i)-
\mathbb E D_{\mathrm{KL}}
\left(
q_i^v\|q_i^r
\right). 
\end{aligned}
\end{equation}
Substituting this relation into \eqref{eq:vrvg_mi_sz} gives
\begin{equation}
\label{eq:vrvg_variational_alignment}
\begin{aligned}
I(\mathbf s_i;\mathbf g_i)
&\ge
h(\mathbf z_i)
+
\mathbb E\log q_{\phi_v}(\mathbf z_i\mid\mathbf g_i)
-
\mathbb E D_{\mathrm{KL}}(q_i^v\|q_i^r)
\\
&=
I(\mathbf g_i;\mathbf z_i)
-
\mathbb E D_{\mathrm{KL}}(q_i^v\|q_i^r).
\end{aligned}
\end{equation}
For a fixed visual embedding distribution $q_{\phi_v}$, the term
$I(\mathbf g_i;\mathbf z_i)$ in
\eqref{eq:vrvg_variational_alignment} is constant, while reducing
$D_{\mathrm{KL}}(q_i^v\|q_i^r)$ increases the corresponding variational
lower bound on $I(\mathbf s_i;\mathbf g_i)$.
Rather than serving as an exact surrogate for mutual information, the KL divergence provides a tractable measure of distributional discrepancy. Minimizing this divergence encourages the radio and visual representations of the same user to exhibit consistent latent semantics, thereby facilitating effective cross-modal grounding.

For balanced alignment of both branches, we exploit a symmetrized KL divergence between radio user $i$ and visual object $j$ with
\begin{equation}
\label{eq:vrvg_distance}
D_{ij}=\frac12\left[
D_{\mathrm{KL}}(q_i^r\|q_j^v)
+D_{\mathrm{KL}}(q_j^v\|q_i^r)\right], 
\end{equation}
where each directed divergence has the closed form \cite{vilnis2015gaussian}
\begin{equation}
\label{eq:vrvg_gaussian_kl}
\begin{aligned}
D_{\mathrm{KL}}(q_i^r\|q_j^v)
=\frac12\sum_{d=1}^{D_z}\bigg[
&\log\frac{(\sigma_{j,d}^v)^2}{(\sigma_{i,d}^r)^2}-1\\
&+\frac{(\sigma_{i,d}^r)^2+
(\mu_{i,d}^r-\mu_{j,d}^v)^2}{(\sigma_{j,d}^v)^2}
\bigg].
\end{aligned}
\end{equation}
The closed form avoids latent sampling and does not require a reconstruction decoder.
Combine same-user alignment with separation from mismatched pairs, we define the VRVG loss as follows\cite{vilnis2015gaussian} 
\begin{equation}
\label{eq:vrvg_loss}
\begin{aligned}
\mathcal L_{\mathrm{VRVG}}
&=\underbrace{\frac1U\sum_{i=1}^{U}D_{ii}}_{\text{same-user grounding}}\\
&\quad+
\underbrace{\frac{1}{U(U-1)}
\sum_{i=1}^{U}\sum_{j\ne i}
[m+D_{ii}-D_{ij}]_+}_{\text{cross-user separation}},
\end{aligned}
\end{equation}
where $m>0$ is the margin and $[x]_+=\max(x,0)$. 
Such grounding and relative separation thus encourage spatially informative, user-discriminative radio representations. 

VRVG is optimized jointly with the subsequent task losses. The variational heads are needed only during training; inference uses the learned deterministic radio tokens without ground-truth boxes or alignment step.

\begin{figure*}[t]
\centering
\includegraphics[width=\textwidth]{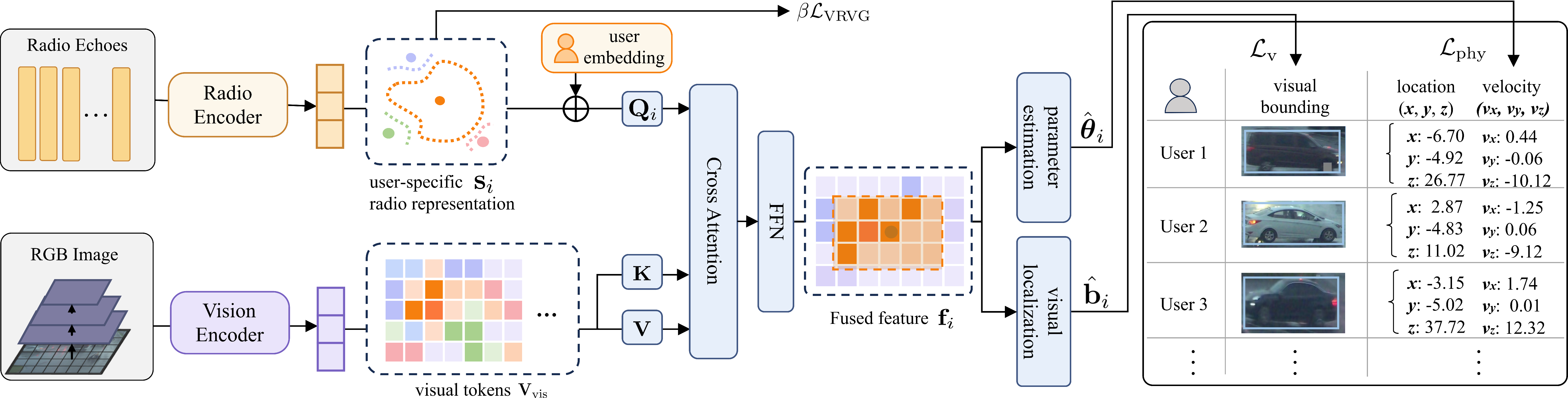}
\caption{Radio-conditioned multimodal fusion network for semantic sensing. 
The user-specific radio representation forms the query for cross-attention, while entangled visual tokens provide the keys and values. The retrieved multimodal features are used to jointly predict the visual bounding box and estimate the user's physical parameters.}
\label{fusion}
\end{figure*}

\subsection{Visual Feature Extraction}
\label{sbsc:pvtsection}

Given the RGB image
$\mathbf V\in\mathbb R^{H\times W\times3}$,
we employ PVTv2 \cite{pvtv2} to extract multi-scale visual features.
For the selected backbone stages $\ell\in\mathcal L_v$, the corresponding
feature maps $\mathbf F_\ell^v$ are flattened and projected with positional encodings $\mathbf E_\ell^v$ 
\begin{equation}
\label{eq:pvt_tokens}
\mathbf T_\ell^v
=
f_{\mathrm{proj},\ell}^v
\!\left(\operatorname{Flatten}(\mathbf F_\ell^v)\right)
+\mathbf E_\ell^v .
\end{equation}
The resulting multi-scale tokens are concatenated and processed by a
Transformer encoder $\operatorname{Enc}_{\mathrm{vis}}$to form the visual
tokens \cite{vaswani2017attention} 
\begin{equation}
\label{eq:pvt_memory}
\mathbf V_{\mathrm{vis}}
=
\operatorname{Enc}_{\mathrm{vis}}
\!\left(
\operatorname{Concat}_{\ell\in\mathcal L_v}\mathbf T_\ell^v
\right)
\in\mathbb R^{L_s\times D}.
\end{equation}
The visual token encodes the spatial structure, and contextual information of users and surrounding objects, while it does not explicitly identify which visual entity corresponds to a particular ISAC user. In the subsequent multimodal fusion, the grounded radio representations serve as user-specific semantic queries to retrieve and refine the corresponding visual features, enabling implicit radio-visual user matching and radio-conditioned visual localization.

\section{Radio-Visual Multimodal Fusion}
\label{sec:multimodal_fusion}
The learned radio representations provide user-specific spatial cues for interpreting the visual scene.
We first describe radio-conditioned visual feature retrieval and then explain the semantic consistency and complementarity underlying the resulting predictions. Finally, we present the joint training objective and strategy.

\subsection{Radio-Conditioned Multimodal Fusion}
As shown in Fig.~\ref{fusion}, the user-specific radio representation $\mathbf s_i$ is augmented with a learned user-index embedding $\mathbf e_i\in\mathbb R^{1\times D}$ to form the semantic query $\mathbf q_i=\mathbf s_i+\mathbf e_i\in\mathbb R^{1\times D}$. 
The radio queries interact with the visual tokens $\mathbf V_{\mathrm{vis}}\in\mathbb R^{L_s\times D}$ through multi-head cross-attention \cite{vaswani2017attention}.
For the $h$-th attention head,
\begin{equation}
\label{eq:crossattn}
\begin{aligned}
\mathbf H_{i,h}^{c}
&=\operatorname{Softmax}\left(
{\mathbf Q^c_{i,h} {\mathbf K_h^c}^T}/{\sqrt{d^c}}\right)\mathbf V^c_h,\\
\mathbf c_i
&=\operatorname{Concat}_{h=1}^{N^c}
(\mathbf H_{i,h}^{c})\mathbf W^{c},
\end{aligned}
\end{equation}
where $\mathbf Q^c_{i,h}=\mathbf q_i\mathbf W_{Q,h}^{c}$, $\mathbf K^c_h=\mathbf V_{\mathrm{vis}}\mathbf W_{K,h}^{c}$, and $\mathbf V_h=\mathbf V_{\mathrm{vis}}\mathbf W_{V,h}^{c}$ are the query, key, and value matrices, respectively.
Here, $N^c$ denotes the number of heads, $d^c=D/N^c$, and the $\mathbf W^{c}$ matrices are learnable attention projections.
Residual connections and an FFN then applied for the final fused feature
\begin{equation}
\label{eq:fused_token}
\begin{aligned}
\mathbf f_i=\operatorname{LN}\bigl(
\operatorname{LN}(\mathbf q_i+\mathbf c_i)+\operatorname{FFN}(\mathbf q_i+\mathbf c_i)\bigr)
\in\mathbb R^{1\times D}.
\end{aligned}
\end{equation}
The proposed fusion pipeline thus transforms unordered visual tokens into ordered user-specific
multimodal semantics according to radio echoes.

Each fused feature is then mapped to the user-specific visual bounding
and physical parameters via two prediction heads 
\begin{equation}
\hat{\mathbf b}_i=f_{\mathrm{box}}(\mathbf f_i)\in[0,1]^4, \quad  
\hat{\boldsymbol\theta}_i=f_{\mathrm{est}}(\mathbf f_i)\in\mathbb R^6.
\end{equation}
Here, $f_{\mathrm{box}}$ predicts the normalized bounding box center, while $f_{\mathrm{est}}$ estimates the user's three-dimensional position and velocity.
Each output remains indexed by its associated pilot, and its correspondence with the visual entity is learned through user-aligned supervision.
Thus, the network jointly performs radio--visual user matching and visual localization through feature-level interaction.

\subsection{Radio-Visual Semantic Consistency and Complementarity}
\label{sec:fusion_insight}

The feature interactions in Fig.~\ref{fusion} illustrate how the learned
radio representations support visual inference.
In the upper path, the highlighted user and surrounding reflectors represent
the spatial context contained in the superimposed echoes.
The user-specific radio representations therefore encodes not only the intended
user's reflection, but also visual task-relevant spatial relationships with other
users and non-user reflectors. 
During training, VRVG encourages this context-aware radio representation to
be grounded to the corresponding visual geometry for identifying and localizing the visual entity.

In the lower path, the visual features contain appearance and spatial
evidence from multiple objects, which can lead to feature
entanglement in crowded scenes \cite{feature_entanglement}.
The radio representations provide user-specific spatial conditioning to the
cross-modal fusion, emphasizing visual features that are consistent
with the radio context of the intended user while suppressing less relevant
components, thereby facilitating the refinement of the corresponding visual region.

The user-specific fused token subsequently produces the visual and
physical outputs shown on the right of Fig.~\ref{fusion}.
Since both predictions are derived from the shared multimodal
representation, visual context can also complement the radio-derived
spatial and motion cues in physical-parameter estimation.
The proposed framework therefore realizes both radio--visual user matching and radio-conditioned multimodal inference.

\subsection{Joint Training Loss}
\label{sec:joint_training}
To train these feature interactions, we couple user-specific localization and the VRVG representation alignment in Section~\ref{sec:vrvg}.
For each scene, the $i$-th prediction is supervised by the box and physical parameter of ISAC user $i$.

We adopt the standard combination of box regression and GIoU losses for visual localization\cite{giou}. Let $\mathcal L_1$ denote the mean absolute error of the normalized box coordinates over all users and the four center-size components, and define
\begin{equation}
\label{eq:giou_loss}
\mathcal L_{\mathrm{GIoU}}
=\frac1U\sum_{i=1}^{U}
\left[1-\operatorname{GIoU}(\hat{\mathbf b}_i,\mathbf b_i)\right].
\end{equation}
The vision localization loss is thus given by
\begin{equation}
\label{losssem}
\mathcal L_{\mathrm{v}}
=\lambda_1\mathcal L_1+\lambda_g\mathcal L_{\mathrm{GIoU}},
\end{equation}
where $\lambda_1$ and $\lambda_g$ balance coordinate regression and region overlap.
User-specific supervision matches each prediction with its visual entity and refines the visual boundary.

We employ mean-squared error (MSE) for position and velocity parameter regression,
denoted by $\mathcal L_{\mathrm{loc}}$ and $\mathcal L_{\mathrm{vel}}$,
respectively. The physical-parameter loss is thus defined as $
\mathcal L_{\mathrm{phy}}
=
\lambda_l\mathcal L_{\mathrm{loc}}
+
\lambda_v\mathcal L_{\mathrm{vel}},
$ where $\lambda_l$ and $\lambda_v$ control the contributions of position
and velocity regression, respectively. The complete training objective is
then given by 
\begin{equation}
\label{ltotal}
\mathcal L_{\mathrm{total}}
=
\mathcal L_{\mathrm{v}}
+
\mathcal L_{\mathrm{phy}}
+
\beta\mathcal L_{\mathrm{VRVG}},
\end{equation}
where $\beta$ controls the contribution of the VRVG loss in
\eqref{eq:vrvg_loss}. The task losses supervise the final predictions
through the multimodal fusion network, whereas VRVG directly grounds the
pre-fusion radio representation in the spatial semantics of its
corresponding visual entity. Jointly, these objectives encourage the radio
branch to learn user-specific semantics that provide guidance for visual localization while preserving the information required for physical-parameter estimation.

% \begin{algorithm}[t]
% \caption{Two-Stage Training of the Proposed Framework}
% \label{alg:training}
% \begin{algorithmic}[1]
% \REQUIRE Paired training data $\mathcal D$; stage lengths $E_1$ and $E-E_1$.
% \ENSURE Trained framework parameters $\boldsymbol\xi$.
% \STATE Initialize the model and load encoder checkpoints.
% \FOR{stage $s\in\{\mathrm{I},\mathrm{II}\}$}
%     \STATE Freeze/unfreeze the visual encoder for Stage I/II.
%     \STATE Reset AdamW and the stage-wise cosine schedule.
%     \FOR{each epoch and minibatch in stage $s$}
%         \STATE Encode inputs and form geometry descriptors $\{\mathbf g_i\}$.
%         \STATE Predict $\{\hat{\mathbf b}_i,\hat{\boldsymbol\theta}_i\}$ and compute $\mathcal L_{\mathrm{VRVG}}$.
%         \STATE Update unfrozen parameters using \eqref{ltotal}, averaged over sensing conditions.
%     \ENDFOR
% \ENDFOR
% \end{algorithmic}
% \end{algorithm}

\section{Numerical Results}
In this section, we evaluate the proposed framework under practical sensing scenarios and present numerical results to validate its effectiveness.

\subsection{RCooper-Based Multimodal Sensing Scenario}
\label{subsec:dataset_settings}

\begin{figure}[t]
\centering
\includegraphics[width=3.3in]{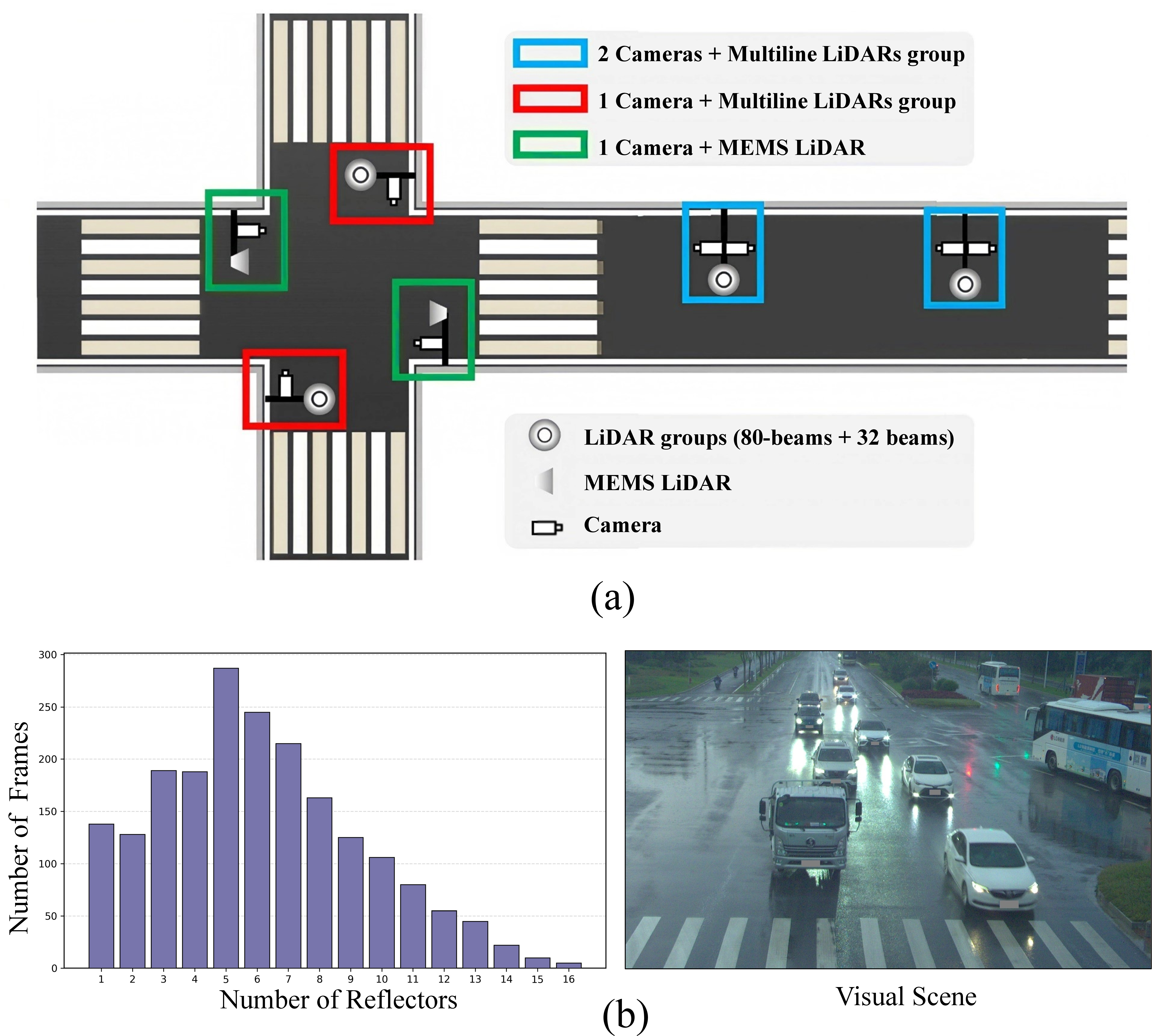}
\caption{RCooper roadside sensing scenes dataset: 
(a) Sensor deployment in the selected roadside scenes. 
(b) Distribution of the number of modeled reflectors together with a representative RGB image. The main experiments use the intersection scenes containing relatively dense reflectors.}
\label{rcooperfig}
\end{figure}

To validate the proposed framework in practical environments, we employ the RCooper dataset \cite{RCooper} as illustrated in Fig.~\ref{rcooperfig}, a large-scale roadside cooperative perception dataset collected in real traffic environments using synchronized cameras and LiDAR sensors, to construct consistent OFDM and visual observations. RCooper provides calibrated camera data together with LiDAR-based 3D object position annotations. 
The objects in each scene are instantiated as OFDM radio reflectors. A subset of them is served as the ISAC users, while the remaining are retained as non-user reflectors. For the pilot beam of a specific user, the simulator computes the propagation delay, Doppler shift, and array responses of each reflector according to its LiDAR-annotated spatial position and its velocity estimated from the frame-to-frame displacement. The received OFDM pilot signal is then generated by superimposing the echoes from the intended user, the other served users, and the non-user reflectors. Consequently, the radio encoder is required to extract user-specific radio representations directly from the multi-object superimposed OFDM echoes. Table~\ref{tab:system_params} summarizes the default OFDM signal configurations. 
For the visual branch, each camera frame is paired with the OFDM sensing observation generated from the same scene geometry. Each image is resized to $H\times W=768\times480$. The valid frames are divided into training, validation, and test sets with a ratio of $80\%$, $10\%$, and $10\%$, respectively.

\begin{table}[t]
\caption{OFDM Signal Parameters}
\label{tab:system_params}
\centering
\footnotesize
\setlength{\tabcolsep}{3pt}
\renewcommand{\arraystretch}{1.1}
\begin{tabular}{@{}p{0.43\columnwidth}p{0.22\columnwidth}p{0.28\columnwidth}@{}}
\toprule
\textbf{Parameter} & \textbf{Symbol} & \textbf{Value} \\
\midrule
Carrier frequency & $f_c$ & 30 GHz \\
Subcarrier spacing & $\Delta f$ & 120 kHz \\
Symbol interval & $T_{\mathrm{sym}}$ & $8.33~\mu$s \\
Tx UPA & $N_t$ & $2\times2;\,16\times16$ \\
Rx UPA & $N_r$ & $1\times1;\,8\times8$ \\
Subcarriers; symbols & $N_s;\,T_s$ & $8:128;\,8:128$ \\
Transmit power & $P$ & $-20:10:40$ dBm \\
Noise power& $\mathbf{z}_i$ & -60 dBm \\
\bottomrule
\end{tabular}
\end{table}

\subsection{Network and Parameter Settings}
\label{subsec:network_settings}
\begin{figure}[t]
\centering
\includegraphics[width=3in]{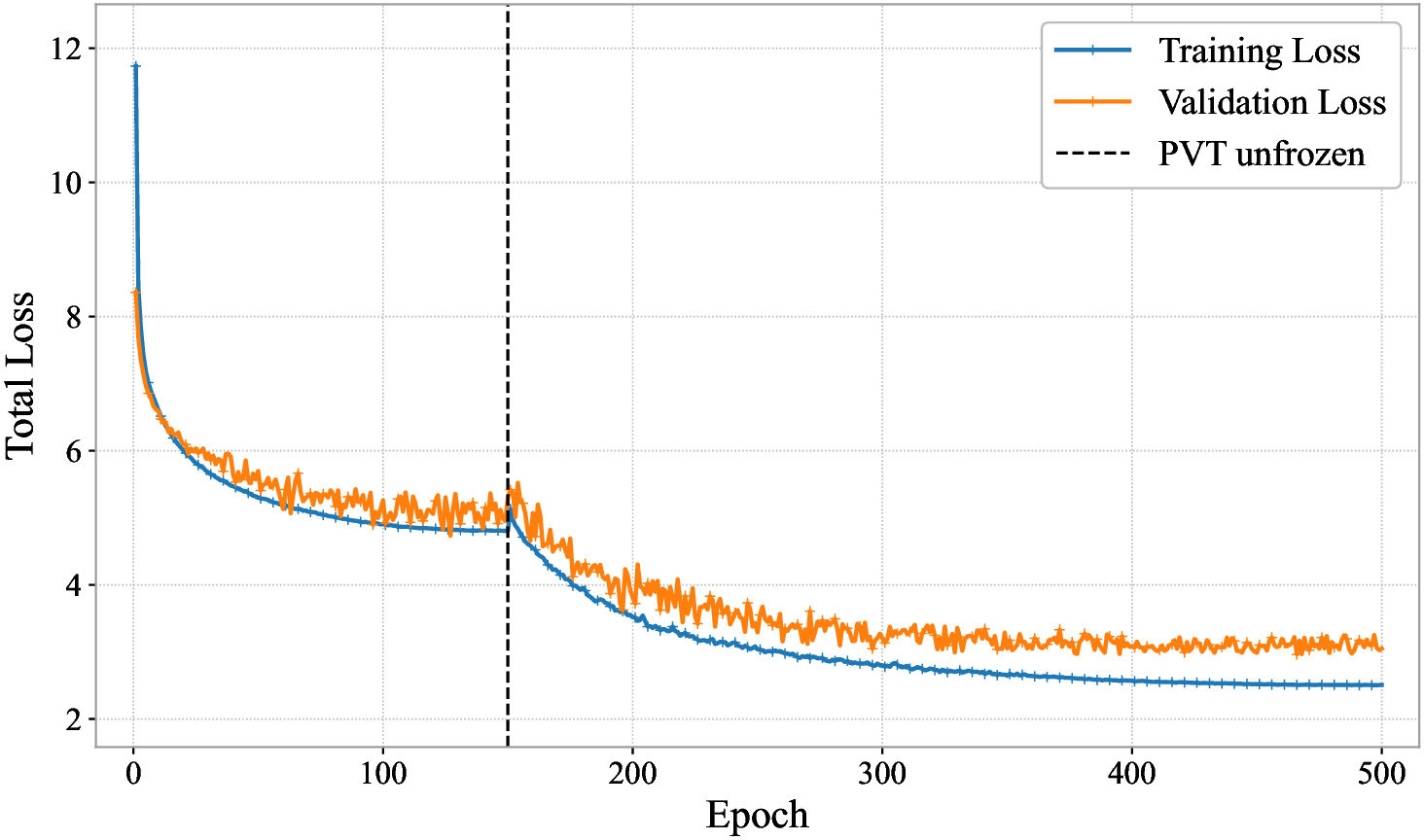}
\caption{Training and validation loss curves under the two-stage strategy.}
\label{traincruve}
\end{figure}

The radio encoder uses 1 complex convolution layer with 64 complex output channels, followed by gated compression with $K=12$ maps and softmax temperature $\tau=0.5$.
The attention-map channels are $48\rightarrow 24\rightarrow12$. 
VRVG uses latent dimension 32 and hidden width 128.
For visual branch, publicly available PVTv2-B2 \cite{pvtv2} supplies the last two feature scales to a 2 layer, 8 head visual Transformer, yielding $L_s=180$ tokens.
Radio self-attention and cross-attention each use 8 head layer. 
The shared feature dimension is 128. 
The weights $(\lambda_1,\lambda_g,\lambda_p,\lambda_v,\beta)$ are $(7,4,10,5,1)$. 
AdamW uses learning rate $10^{-4}$, weight decay $10^{-4}$, and cosine annealing to $10^{-6}$, with the optimizer and schedule reset at the stage transition.
As illustrated in Fig~\ref{traincruve}, during the training process, we fix the ImageNet-pretrained PVTv2-B2 visual encoder for 150 epochs to provide a stable visual feature basis for radio representation learning, and jointly training them for another 350 epochs. 
The variational heads remain trainable in both stages, and we reconfigure AdamW and its stage-specific learning-rate schedule at the stage transition. 
At inference, the VRVG supervision and variational heads are omitted, and the main path in Fig.~\ref{fusion} uses only the observed radio echoes and RGB image.

\subsection{Baseline Methods for Comparison}
\label{subsec:baseline_methods}

We compare radio compression, alignment, visual-only, and physical estimation-based user matching to assess where radio representations benefit multimodal inference.

\noindent(1) \emph{CNN Compression:}
This baseline replaces our gated attention compression module with separate CNN branches and full connect layers for the real and imaginary echo components\cite{ofdmDL3}. 
The network is designed to estimate the positions and velocities of multiple point-like targets from OFDM echoes, while it does not explicitly account for other reflectors.
In our implementation, each branch uses two 3-D convolutional layers with 32/64 channels and $4^3$ kernels, followed by global average pooling; the concatenated branch features are projected to 128 dimensions.

\noindent(2) \emph{CConv Compression:}
This baseline replaces our gated attention compression module with  3 layers of complex-valued convolution (CCconv). 
The network is designed for LFM radar echoes to identify one ISAC user among multiple targets and estimate its physical parameters, while ignoring echoes from other targets  \cite{SIMAC}.
Our implementation uses 16/32/64 complex channels with $3^3$ kernels; two $2^3$ max-pooling stages and final global max pooling yield 128-dimensional real-valued tokens.

\noindent(3) \emph{No MHSA and No VRVG:}
Removing radio MHSA to test the multi-user context learning; Removing VRVG loss to test the contribution of VRVG alignment.

\noindent(4) \emph{InfoNCE:}
This baseline method employs a multimodal contrastive alignment in place of VRVG \cite{radford2021clip}. While it utilizes the same radio features and visual descriptors as VRVG, it replaces variational grounding with an InfoNCE loss based on the cosine similarity of projected features.
Both projection MLPs use widths $128\rightarrow128\rightarrow64$.

\noindent(5) \emph{Oracle Vision:}
We evaluate the visual-only approach under an Oracle Vision setting. Ground-truth user identity is used to establish a one-to-one correspondence between each served user and its maximum IoU prediction. We further exclude confidence filtering and false-positive penalties, giving the visual detector a deliberately favorable evaluation condition. Consequently, the comparison can verify whether radio representations can further refine the localization of the already identified visual entity. 
The detector uses 128-dimensional features and a two-layer, four-head Transformer decoder with FFN width 512.

\noindent(6) \emph{Signal-only and FFT estimators:}
The \emph{Signal-only} estimator uses our radio signal branch with a $128\rightarrow128\rightarrow6$ regression MLP to infer position and velocity parameters without vision.
Traditional FFT estimator extracts delay, Doppler, and arrival angle from the OFDM echoes to recover position and velocity parameters. 

\noindent(7) \emph{User matching:}
Following post-alignment and matching paradigm, we combine physical estimators in (6) with DNN or Hungarian matching, yielding four output-level baselines.
The DNN scores each estimated radio state against each visual candidate using physical estimations and selects the highest-scoring candidate per user \cite{userIdentification,radarToUserAssociation}.
The matching MLP has widths $128\rightarrow128\rightarrow1$.  
Hungarian matching instead minimizes projected-center distances subject to one-to-one assignment \cite{detr}. 
All four methods use the selected $U$ ground-truth candidate boxes, so their comparison isolates user mathching from visual localization errors.

\subsection{Semantic Localization Performance}

\subsubsection{User Matching Performance}
\label{subsec:user_object_matching_results}

We first evaluate whether each user-specific radio observation is matched to
the correct visual entity. For the prediction of the $i$-th ISAC user, we
compare its predicted $\hat{\mathbf b}_{i}$ with the ground-truth boxes
of all users in the same frame and assign its visual identity to the
candidate with the largest IoU. The match is regarded as correct if this
maximum-IoU candidate is the ground-truth visual entity of user $i$.
The matching accuracy is then computed as the
fraction of correctly matched radio-vision user pairs over all evaluated users.
For the output-level matching baselines, $U$ ground-truth candidate boxes are directly provided for mathching. Therefore, these baselines are evaluated under a favorable matching-only condition. 

\begin{figure}[t]
\centering
\subfloat[\label{fig:matching_accuracy_users}]{%
\includegraphics[width=0.49\columnwidth]{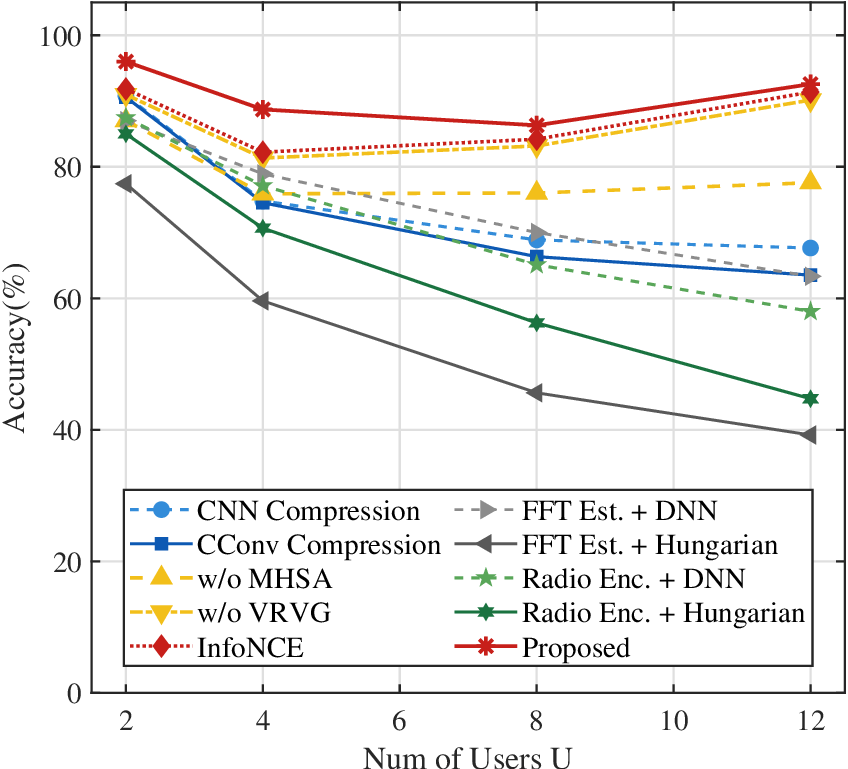}}\hspace{1pt}
\subfloat[\label{fig:matching_accuracy_signal_level}]{%
\includegraphics[width=0.49\columnwidth]{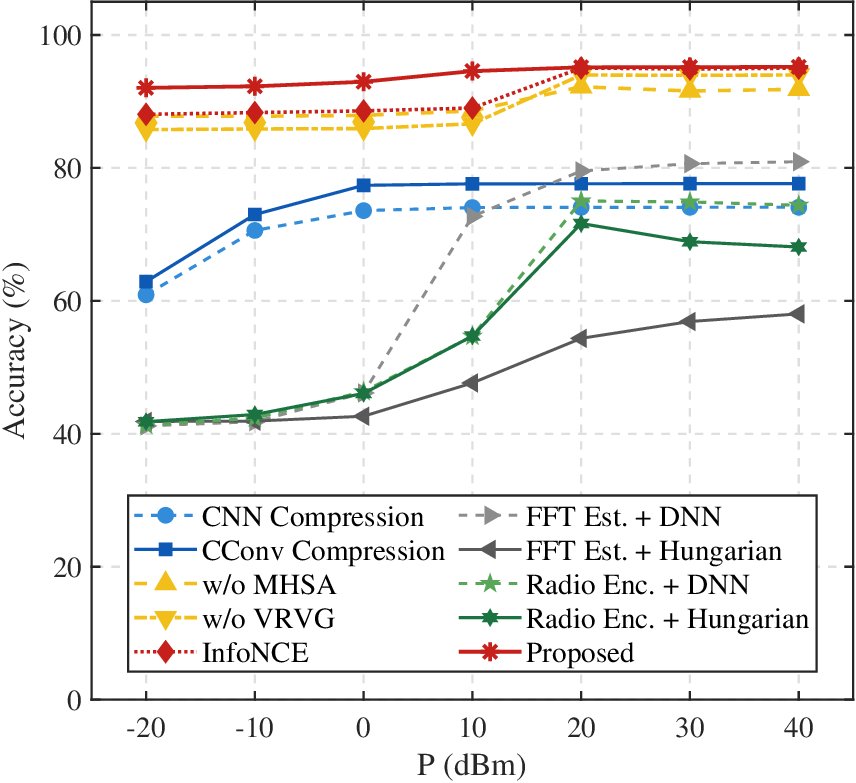}}
\caption{Radio--visual user-matching accuracy under 
(a) different numbers of served ISAC users $U$ and 
(b) different radio transmit power settings P(dBm). 
Accuracy measures whether each radio representation is successfully matched with the correct visual entity.}
\label{fig:user_object_matching}
\end{figure}

Fig.~\ref{fig:matching_accuracy_users} shows the matching accuracy for
different numbers of served users with $U\in\{2,4,8,12\}$.
The proposed framework achieves the
highest accuracy throughout the evaluated settings and remains above
approximately $86\%$. At $U=12$, it achieves approximately $93\%$,
whereas FFT Est.~+~DNN and Radio Enc.~+~DNN achieve approximately
$64\%$ and $58\%$, respectively. The corresponding Hungarian methods
decrease further to approximately $39\%$ and $45\%$.
These results indicate the limitation of the decoupled
physical estimation then radio-vision matching paradigm as the number of candidate
users increases. In contrast, the learned radio representation preserves
visual task-relevant spatial information for direct radio--vision correspondence.
The nonmonotonic trend across \(U\) reflects two competing effects: As more users are introduced, the evaluated scenes generally contain more reflector, leading to stronger echo superposition and making user-specific matching more difficult. Meanwhile, a larger \(U\) also provides additional user-specific pilot observations over different sensing directions, yielding richer spatial information. At larger \(U\), the proposed framework can increasingly exploit such multi-user spatial cues through mutual context learning, which explains the recovery in matching accuracy. In contrast, the improvement is much weaker without MHSA, further supporting the role of multi-user learning in aggregating complementary spatial information across users.

Fig.~\ref{fig:matching_accuracy_signal_level} examines the matching sensitivity to radio transmit power. At the lowest signal level, the proposed method retains approximately $92\%$ matching accuracy, compared with approximately $61\%$ and $63\%$ for CNN and CConv Compression,
respectively, while the physical-estimation pipelines remain much
lower. As $P$ increases, the proposed method approaches $95$--$96\%$,
whereas the output-level matching baseline, FFT Est.~+~DNN saturates
near $81\%$. The persistent gap suggests that learning
visual task-oriented radio semantics directly from the echoes provides
more effective cues for cross-modal correspondence than first estimate the low-level physical parameters. 
The alignment baselines further clarify the role of VRVG. Under degraded
radio observations, removing VRVG or replacing it with InfoNCE reduces
the matching accuracy. This indicates that VRVG can help preserve user-specific semantic consistency when the radio observations are less reliable. 
These results establish the matching benefit; the following analysis separately examines whether the visual boundary of the matched users are also improved.

\subsubsection{Evaluation Metric}
\label{subsec:performance_metrics}

The matching accuracy establishes whether user-specific radio representation identifies the corresponding visual entity. We next evaluate the visual localization task performance.  
Conventional object detection evaluates an unordered set of predictions
using metrics such as mean Average Precision (mAP) and mean Recall (mR),
which involve prediction--ground-truth assignment and confidence ranking \cite{voc2010}. In this semantic sensing framwork, however, 
each ISAC user defines an indexed and is associated with one corresponding visual localization. This radio-conditioned setting naturally reduces localization evaluation to
whether the visual region of with each user sufficiently overlaps its ground-truth.
Therefore, for the ground-truth $\mathbf b_{i}$ and predicted $\hat{\mathbf b}_{i}$, the success rate at IoU threshold $k$ is given by 
\begin{equation}
\label{eq:semantic_success_complexity}
\mathrm{SR}@k
=
\frac{1}{U}
\sum^U_{i=1}
\mathbb I\!\left[
\operatorname{IoU}
(\hat{\mathbf b}_{i},\mathbf b_{i})\geq k
\right].
\end{equation}
Following the IoU-threshold sweep commonly used in visual detection
evaluation, we report $\mathrm{mSR}$ as the average of
$\mathrm{SR}@k$ over
$k\in\{0.50,0.55,\ldots,0.95\}$ \cite{coco_eval}.
Thus, $\mathrm{mSR}$ retains the visual localization criterion of conventional metrics while adapting it to the one-to-one, user-conditioned
multimodal sensing setting. 
The previously defined Oracle Vision baseline is evaluated using
the same criterion after assigning its visual predictions to the
corresponding users.

\subsubsection{User-Conditioned Visual Localization}
\label{subsec:new_visual_localization}

\begin{figure}[t]
\centering
\includegraphics[width=0.7\columnwidth]{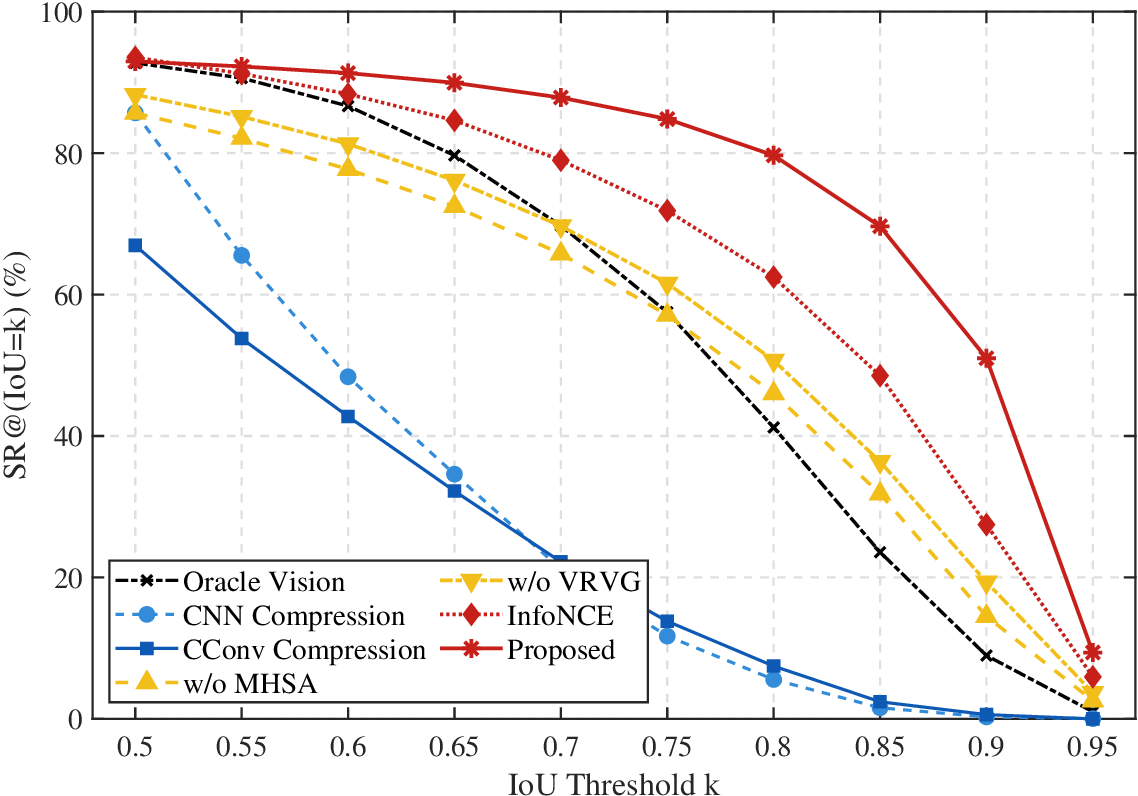}
\caption{Radio-conditioned visual localization success rate $\mathrm{SR}@k$.}
\label{fig:new_sr_iou}
\end{figure}

Fig.~\ref{fig:new_sr_iou} evaluates the visual localization performance using $\mathrm{SR}@k$. At $k=0.5$, the proposed method and
Oracle Vision both achieve over $90\%$ SR, indicating comparable
coarse object localization once the radio-vision matching is established. As the
IoU criterion becomes stricter, however, their separation increases
substantially. At $k=0.9$, the proposed method retains approximately $51\%$
success, compared with $28\%$ for InfoNCE, $19\%$ without VRVG, and only
$9\%$ for Oracle Vision. Since Oracle Vision is already provided with
oracle matching, this gain demonstrates that the learned
radio semantics provide complementary spatial cues for refining spatial extent of the corresponding visual entity, rather than merely achieving user matching. 
Removing MHSA or VRVG causes increasing degradation at high IoU
thresholds. This is consistent with MHSA exploiting the relative spatial
structure across user-specific echo observations, while VRVG
grounds such radio representations toward the spatial structure required
by visual inference. InfoNCE performs better than removing the alignment
objective, but remains clearly below VRVG under stringent visual localization criteria. 
The rapid performance degradation of CNN and CConv compression baselines further indicates that radio encoders designed for physical parameter estimation without accounting for interference are insufficient to support high-level semantic sensing task.

\begin{figure}[t]
\centering
\includegraphics[width=0.96\columnwidth]{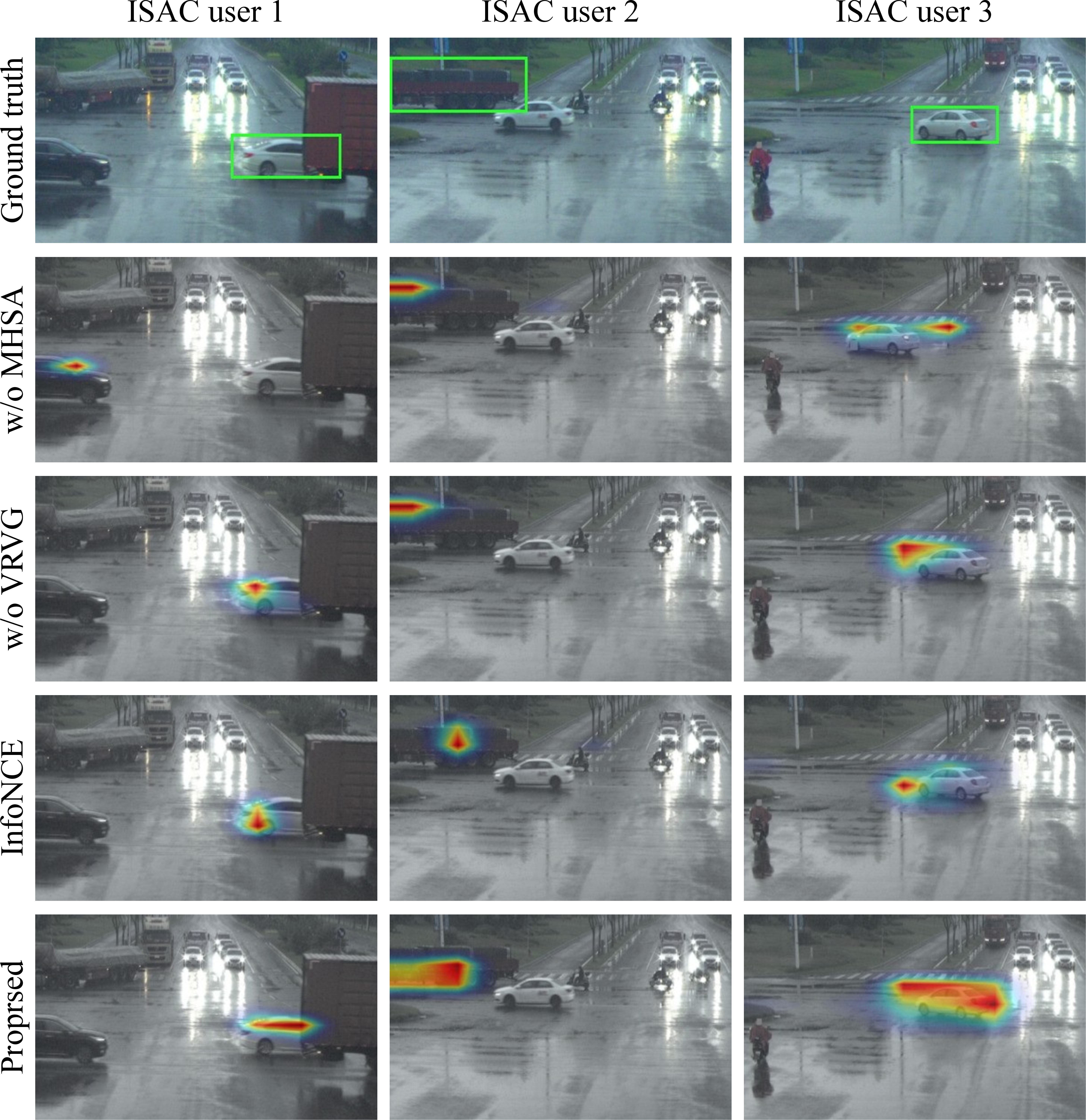}
\caption{Visualization of radio-conditioned cross-attention heatmaps on different users.  
Each column corresponds to a different served ISAC user. 
The visualization illustrates how different user-specific radio queries attend to spatial visual regions.}
\label{fig:vis_attention}
\end{figure}

Fig.~\ref{fig:vis_attention} visualizes the cross-attention heatmaps induced by different user-specific radio queries. Without MHSA, the attention is often shifted or distracted by other reflectors, showing that removing multi-user interaction weakens the ability to exploit relative spatial structure under superimposed echoes and leads to matching error. Without VRVG, the attention becomes more diffuse and less tightly concentrated on the target extent, indicating that radio representations without multimodal grounding are less effective at providing visually aligned spatial cues. InfoNCE introduces cross-modal alignment and can roughly identify the correct object, but its attention mainly concentrates on a single point on the entity, with weaker coverage of the object's full extent. This indicates that this deterministic alignment method fails to effectively learn comprehensive visual task-oriented features, as the features tend to collapse into point-location parameters. In contrast, the proposed method produces more accurate and extended attention on the corresponding visual entity. This result supports that the proposed framework learns both discriminative and vision-oriented radio semantics that are thereby providing more effective guidance for visual localization refinement.

\begin{figure}[t]
\centering
\subfloat[\label{fig:new_msr_users}]{%
\includegraphics[width=0.49\columnwidth]{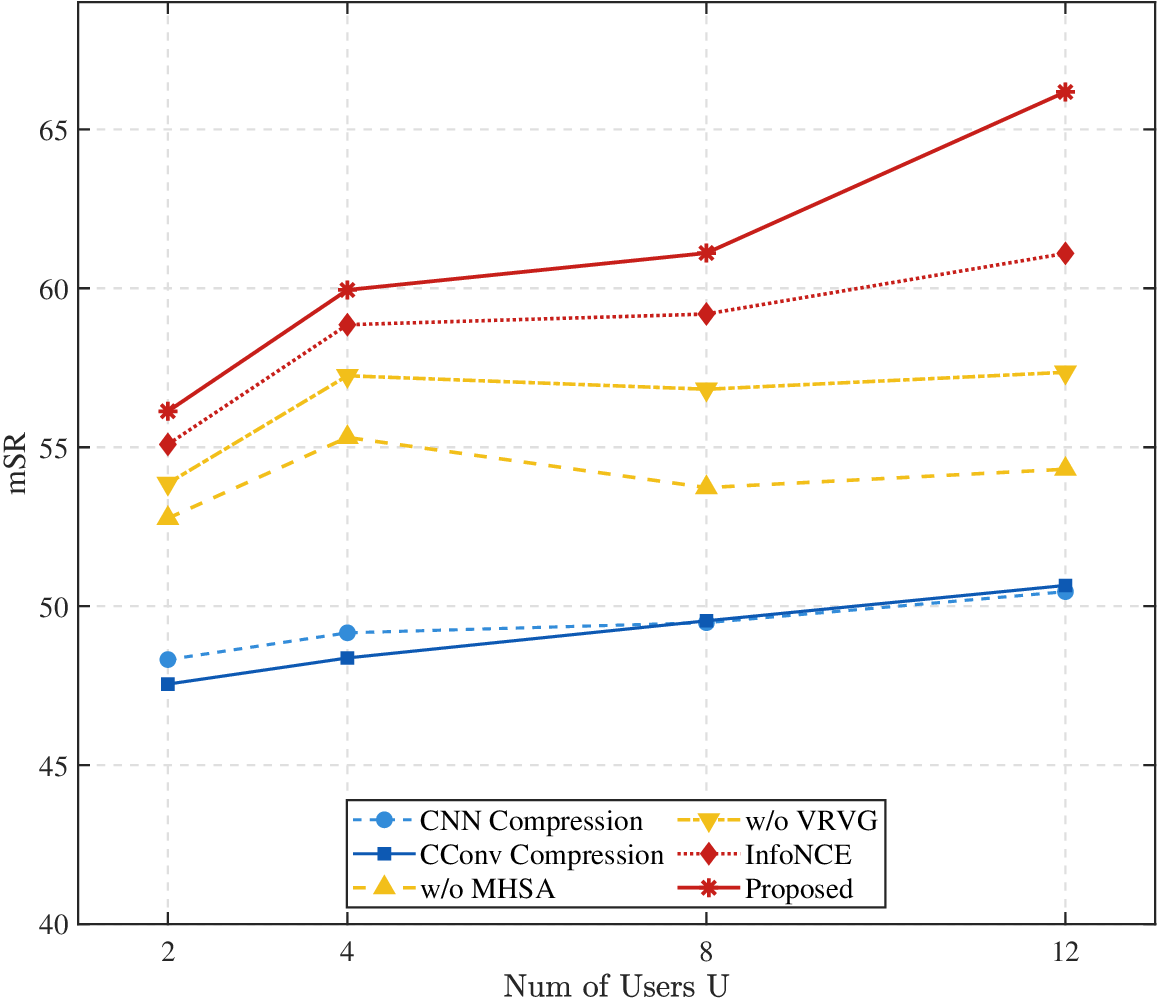}}\hspace{1pt}
\subfloat[\label{fig:new_msr_signal}]{%
\includegraphics[width=0.49\columnwidth]{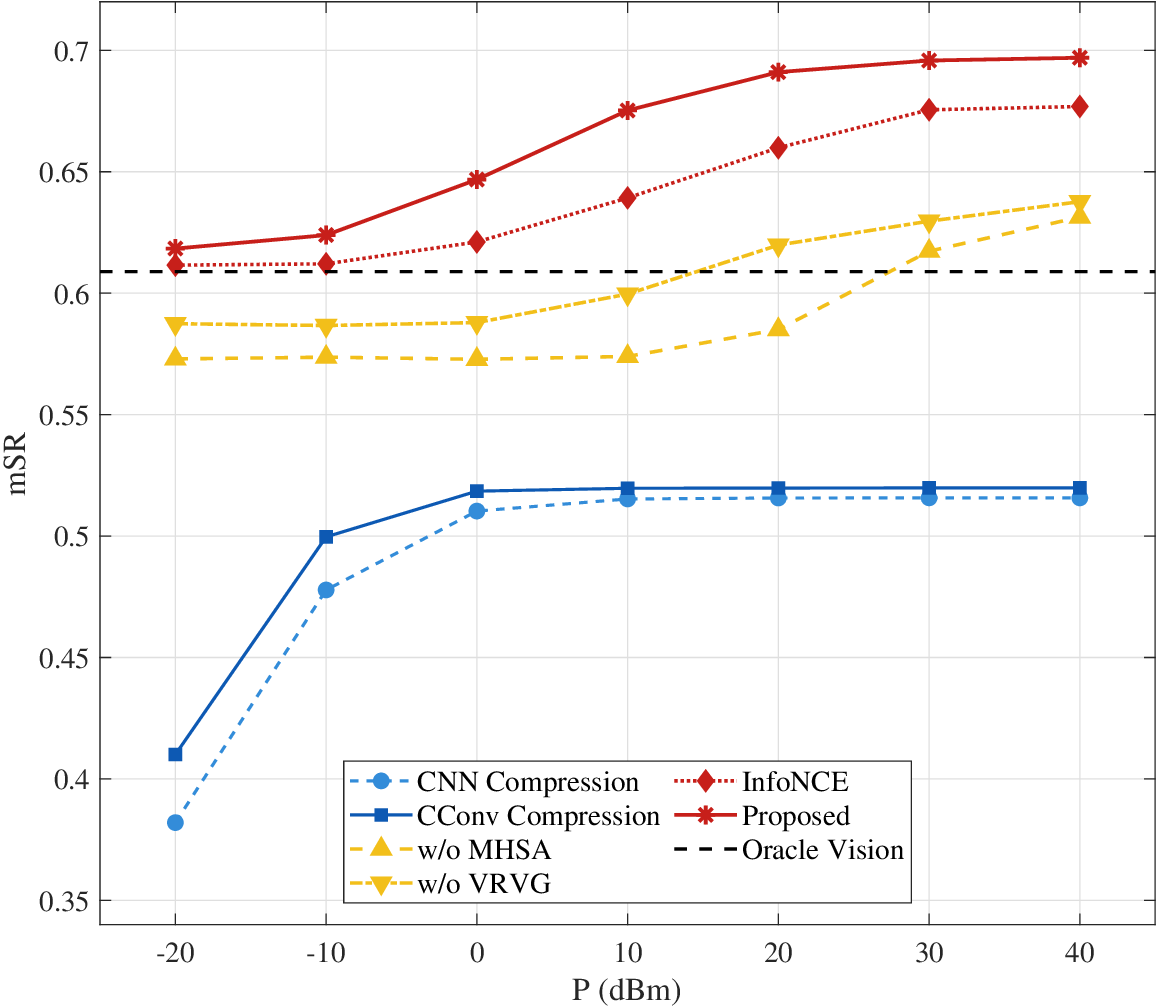}}
\caption{Radio-conditioned visual localization performance measured by $\mathrm{mSR}$ versus 
(a) the number of served ISAC users $U$ and 
(b) the radio signal transmit power P(dBm).}
\label{fig:new_msr_users_signal}
\end{figure}

We next quantitatively evaluate the overall localization performance using
$\mathrm{mSR}$. Fig.~\ref{fig:new_msr_users_signal} summarizes the results under different numbers of served users and radio transmit powers.
As shown in Fig.~\ref{fig:new_msr_users}, the proposed method achieves the
highest $mSR$ for all evaluated $U$. InfoNCE is the closest alternative, whereas
removing MHSA or VRVG consistently degrades localization accuracy.
Increasing $U$ introduces more reflectors and hence stronger echo
superposition, but also activates more available sensing information
from different users, providing richer spatial information.
The proposed framework can exploit these complementary multi-user cues,
leading to a gain at larger $U$. In particular, the small
improvement without MHSA further supports the role of multi-user
context learning in aggregating the relative spatial structure contained in the
superimposed echoes. 

Fig.~\ref{fig:new_msr_signal} further shows how the visual localization gain depends
on radio observation quality. The proposed $mSR$ increases from approximately
$0.6$ at $P=-20$ dBm to nearly $0.70$ at high transmit power, while Oracle
Vision remains around $0.6$. The increasing separation indicates that radio sensing provide useful spatial semantics for refining the corresponding visual entity beyond visual-only inference. At high transmit power, InfoNCE reaches approximately $0.68$,
whereas removing VRVG or MHSA yields only about $0.64$ and $0.63$,
respectively. Together with the SR@\(k\) results, this confirms that
radio--vision alignment is not only beneficial for identifying the correct
visual entity, but also enables the learned radio semantics to provide
complementary spatial features for fine-grained visual localization.

\subsubsection{Multi-Object Echo Interference}
\label{subsec:new_multiobject_results}

\begin{figure}[t]
\centering
\includegraphics[width=0.93\columnwidth]{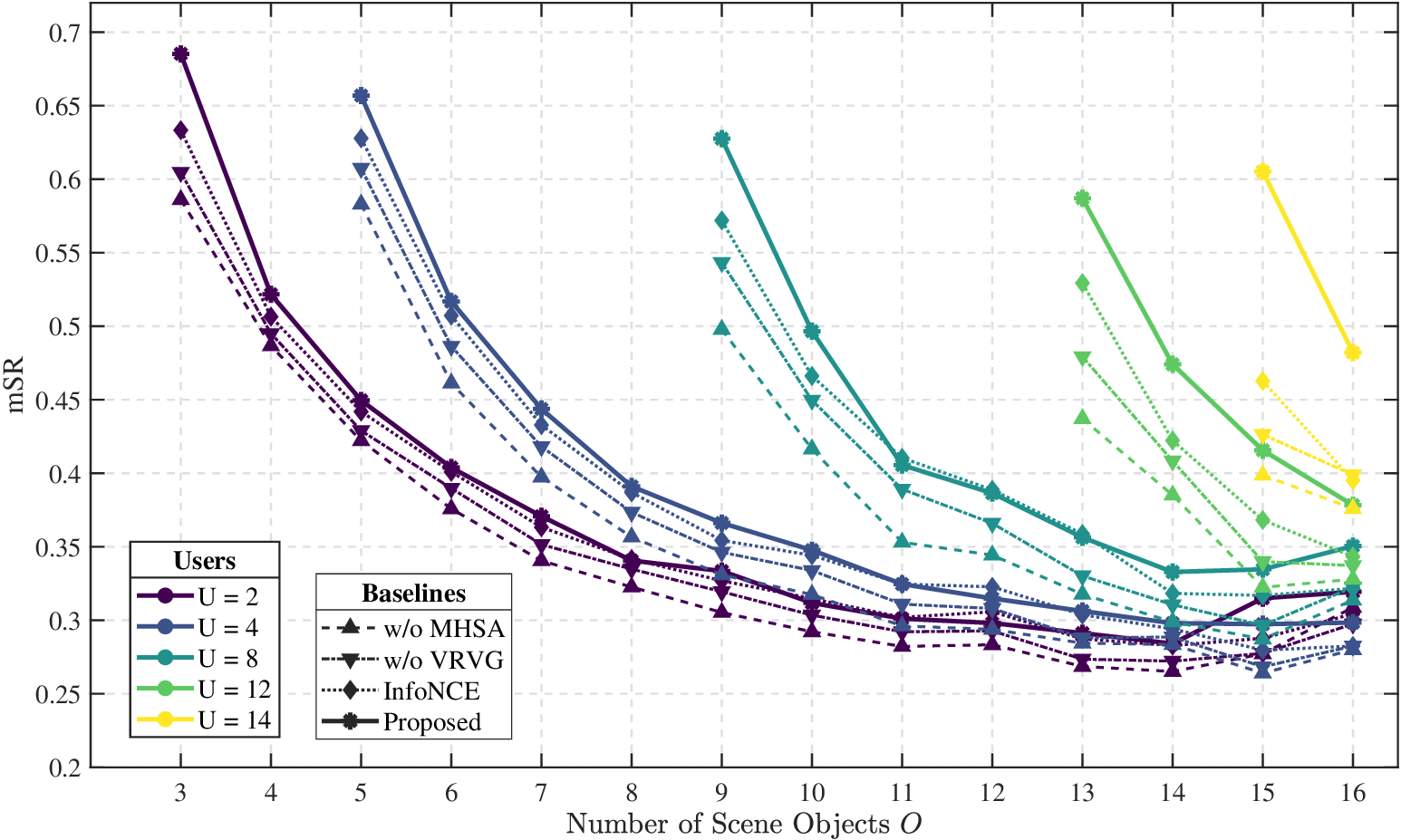}
\caption{$\mathrm{mSR}$ versus the total number of modeled reflectors $O$ for different numbers of served users $U$ with $O\ge U$. For a particular ISAC user, the echoes from the other users and the non-user reflectors are superimposed with the intended-user echo.}
\label{fig:new_msr_objects}
\end{figure}

To further analyze the effects of served users and surrounding reflectors, Fig.~\ref{fig:new_msr_objects} reports the $mSR$ performance jointly with the number of served users $U$ and the total number of modeled reflectors $O$. For a fixed $U$, increasing $O$ corresponds to introducing more non-user reflectors, whose echoes are superimposed with those of the served users. Accordingly, the $mSR$ generally decreases as $O$ increases. For example, the proposed method drops from approximately $0.69$ at $(U,O)=(2,3)$ to about $0.32$ at $(2,16)$. This trend confirms that echoes from additional non-user reflectors introduce substantial ambiguity into user-specific radio representation learning. The complementary effect of $U$ can also be observed when comparing configurations with the same reflectors $O$. A larger number of served users provides more user-specific pilot observations and richer inter-user spatial context. The proposed framework can exploit these additional multi-user cues through interaction learning, which helps preserve localization performance even under dense echo superposition. In contrast, removing MHSA significantly reduces the benefit obtained from larger $U$, supporting its role in aggregating relative spatial relationships across user-specific observations. Across different $(U,O)$ scenario, the proposed method also outperforms the variants without VRVG and with InfoNCE. This indicates that, beyond resolving the intended user from superimposed echoes, VRVG preserves radio semantics that remain useful for refining the corresponding visual entity under increasingly complex scene compositions.

\subsection{Joint Physical Parameter Estimation}
\label{subsec:new_physical_results}

\begin{figure}[t]
\centering
\subfloat[\label{fig:new_position_rmse}]{%
\includegraphics[width=0.48\columnwidth]{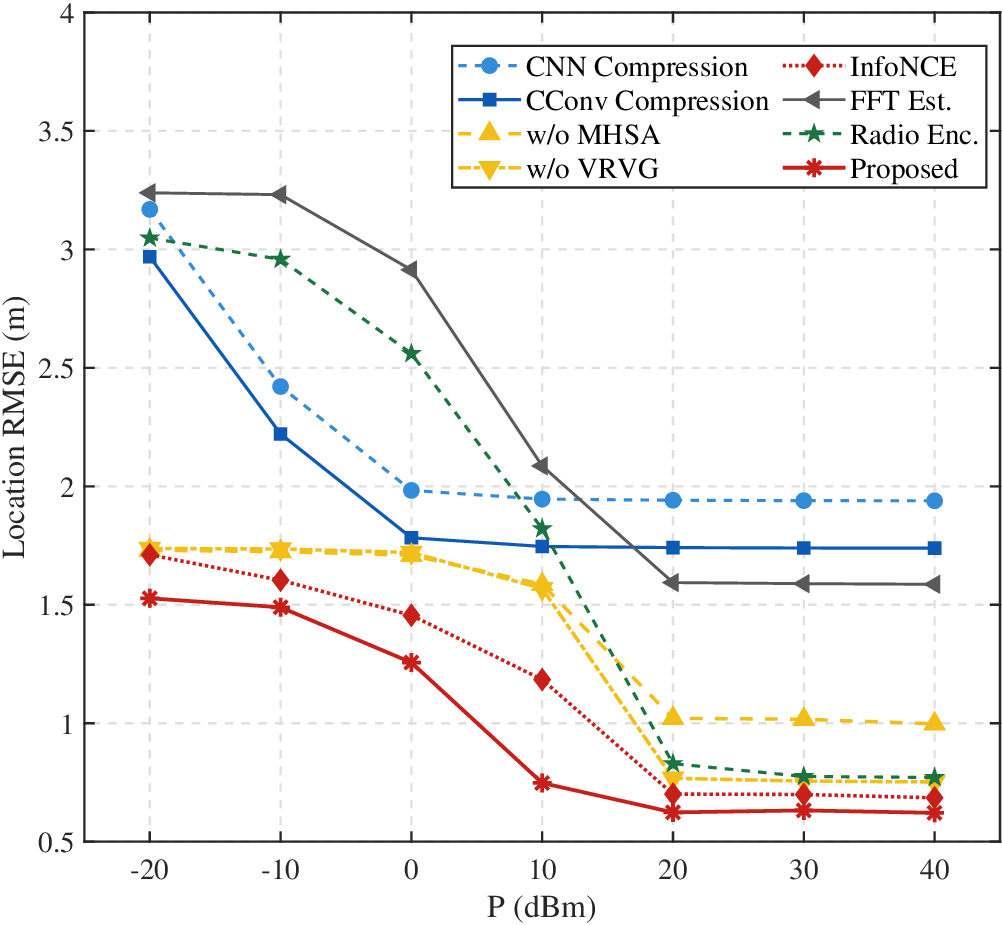}}\hspace{1pt}
\subfloat[\label{fig:new_velocity_rmse}]{%
\includegraphics[width=0.48\columnwidth]{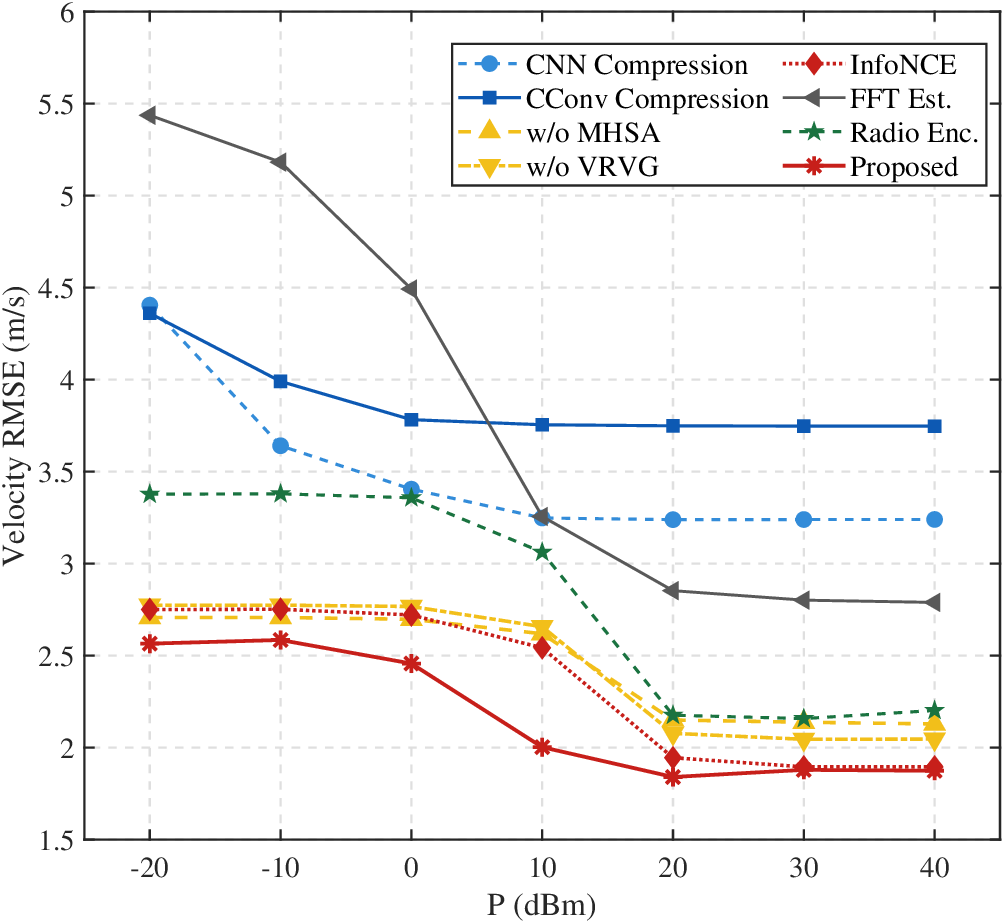}}
\caption{Joint physical parameter estimation under different radio signal transmit power P(dBm): 
(a) position RMSE 
(b) velocity RMSE. }
\label{fig:new_physical_rmse}
\end{figure}

The learned multimodal representation also preserves the information for user-specific physical parameter estimation. As shown in Fig.~\ref{fig:new_physical_rmse}, the proposed method achieves the lowest position and velocity RMSE across the evaluated transmit-power range. Its position RMSE decreases from approximately $1.5$~m at
$P=-20$~dBm to about $0.6$~m at high transmit power, while the velocity
RMSE decreases from approximately $2.6$ to $1.8$~m/s.
In comparison, the signal-only Radio Encoder exhibits substantially larger
errors, particularly under weak radio observations.  This indicates that visual
features provide complementary physical-level information for the user-specific physical state.
Removing VRVG or replacing it with InfoNCE also degrades both position and
velocity estimation, although the gaps become smaller as the radio
observation improves. This suggests that the same radio--vision semantic
consistency that benefits visual localization also helps form a more
user-specific fused representation for physical parameter regression.
The traditional baselines directly recovers delay, angle, and Doppler from the
received echoes. Since Doppler primarily provides radial-velocity
information, reconstructing the full velocity vector introduces an
additional limitation, leading to a larger velocity RMSE.

\subsection{Array and Pilot-Resource Studies}
\label{subsec:new_resource_results}

\begin{figure}[t]
\centering
\subfloat[\label{fig:new_tx_array}]{%
\includegraphics[width=0.49\columnwidth]{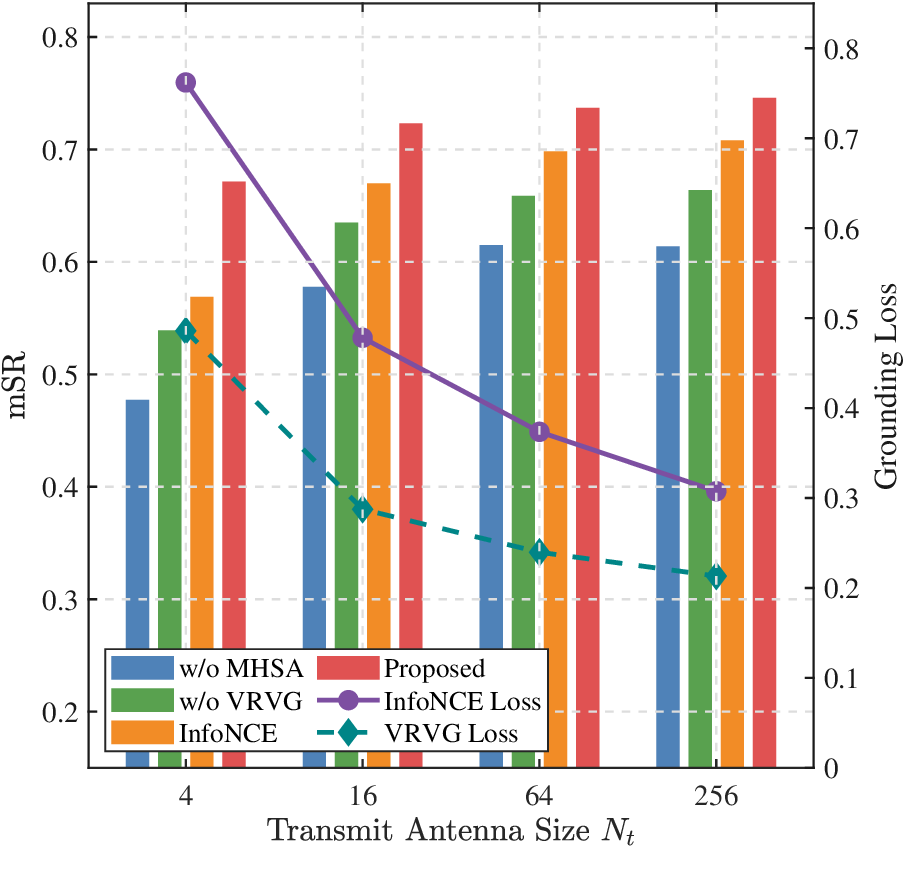}} \hspace{1pt}
\subfloat[\label{fig:new_rx_array}]{%
\includegraphics[width=0.49\columnwidth]{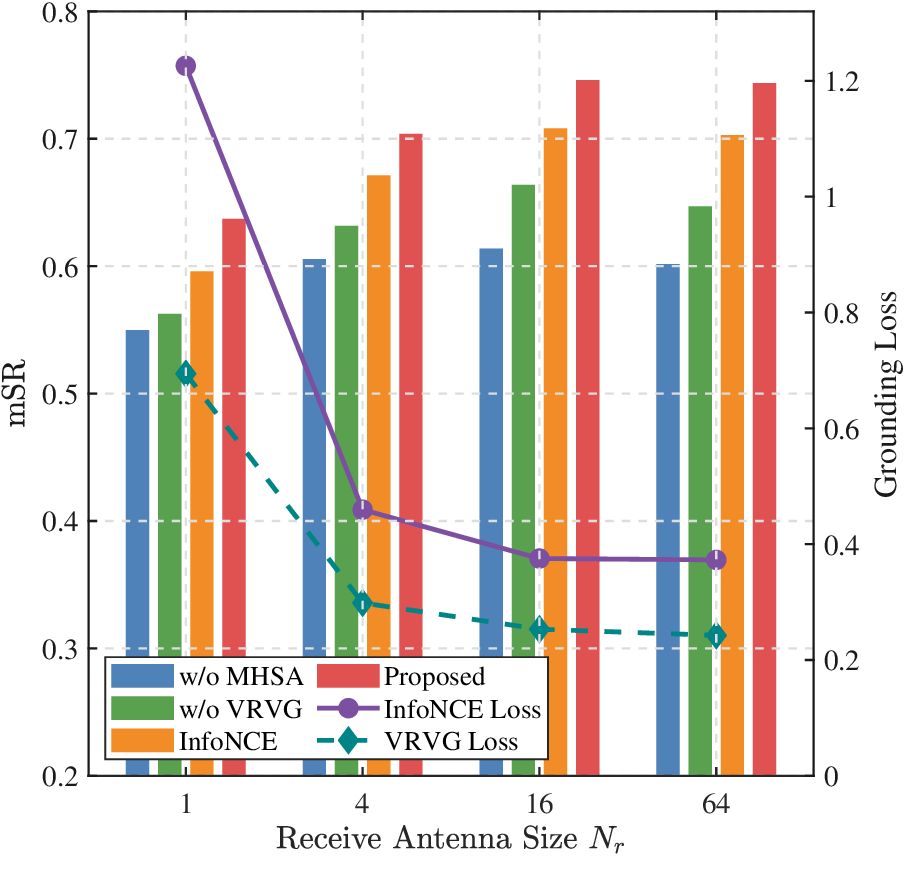}}
\caption{Effects of the numbers of 
(a) transmit antennas $N_t$ and 
(b) receive antennas $N_r$ on visual localization and cross-modal alignment performance. 
The bars show $\mathrm{mSR}$ on the left axis, while the lines show the corresponding alignment losses on the right axis.}
\label{fig:new_antenna_grounding}
\end{figure}

Fig.~\ref{fig:new_antenna_grounding} further studies how spatial sensing resources affect visual localization and radio--vision grounding. As the transmit array $N_t$ increases from $2\times 2$ to $16\times 16$, the proposed $mSR$ increases from approximately $0.67$ to $0.75$. A similar improvement is observed when enlarging the receive array. The proposed method consistently outperforms the variants without MHSA or VRVG and the InfoNCE baseline over all antenna settings. 
The performance improvement can be explained by the richer spatial
observations provided by larger arrays. Increasing $N_t$ improves the
spatial resolution of user-specific sensing, reducing the ambiguity
between the intended user and surrounding reflectors. Increasing $N_r$
enlarges the receive aperture and provides more discriminative angular
responses, which facilitates the separation of superimposed echoes from
different directions. Consequently, the learned radio
representations contain more reliable user-specific and relational spatial
semantics for guiding visual inference.
This effect is also reflected by the alignment losses. For both transmit
and receive array sweeps, the InfoNCE and VRVG losses decrease as the
number of antennas increases. With richer angular observations, the
radio representations of different users become more distinguishable and
their spatial correspondence with the associated visual entities becomes
less ambiguous. Therefore, the decreasing loss indicates that
the learned radio distribution can be grounded more consistently to the
visual spatial representation as the available spatial sensing resource
increases.  
The gains gradually saturate for large arrays, suggesting that once sufficient spatial information has been
captured, further enlarging the array provides limited additional benefit
and the remaining localization error is increasingly determined by the
visual inference and representation-learning stages. The absolute values
of the InfoNCE and VRVG losses are not directly comparable because they
correspond to different optimization objectives; only their trends within
each objective are interpreted here.

\begin{figure}[t]
\centering
\subfloat[\label{fig:new_ns_msr}]{%
\includegraphics[width=1\columnwidth]{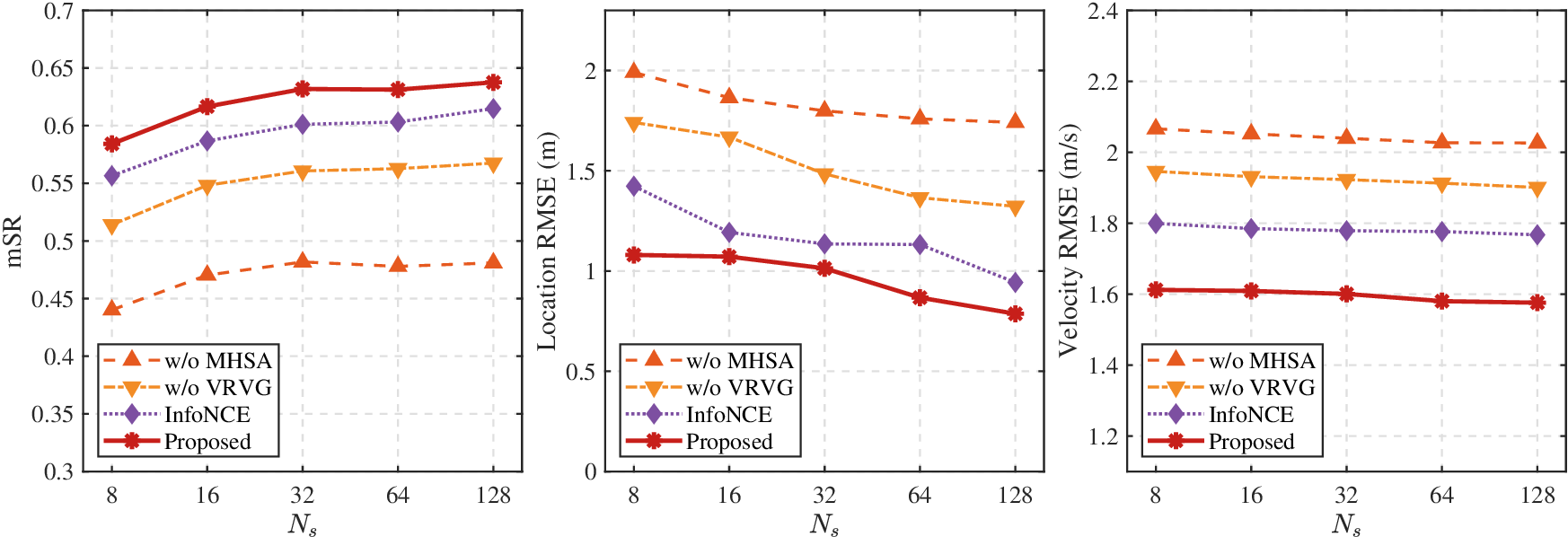}}\\[5pt]
\subfloat[\label{fig:new_ts_msr}]{%
\includegraphics[width=1\columnwidth]{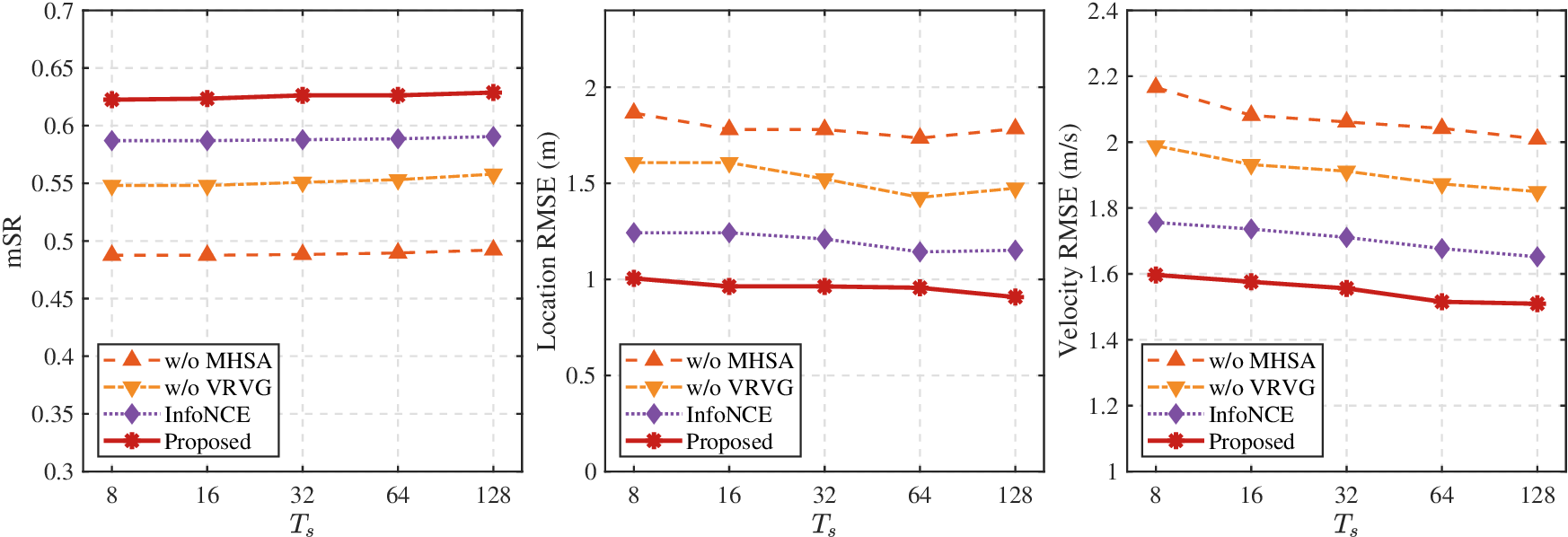}}
\caption{Effects of pilot resources on multimodal semantic sensing performance. 
The top row varies the number of pilot subcarriers $N_s$, while the bottom row varies the number of observed OFDM symbols $T_s$. 
The subfigures report $\mathrm{mSR}$, position RMSE, and velocity RMSE, respectively.}
\label{fig:new_pilot_msr}
\end{figure}

Fig.~\ref{fig:new_pilot_msr} examines how the user-specific pilot resources
affect multimodal semantic sensing and physical parameter estimation. 
As shown in Fig.~\ref{fig:new_pilot_msr}(a), increasing subcarrier counts $N_s$ improves the proposed $\mathrm{mSR}$ from approximately $0.59$ to $0.64$ and reduces the position RMSE from about $1.1$ to $0.8$~m,
whereas the velocity RMSE changes only slightly. More frequency-domain
pilot observations provide finer delay-dependent structure and richer
spatial information in the radio representation, benefiting both position
estimation and radio-conditioned visual localization. The proposed method
consistently achieves the highest $mSR$ and lowest estimation errors over the
entire sweep. Notably, most of the semantic gain is already obtained with
a moderate number of pilot subcarriers, while further increasing $N_s$
occupies additional frequency-domain resource elements that could otherwise
be used for communication payload. This indicates that the learned
visual task-oriented radio representation can provide effective multimodal
inference without requiring a large resource allocation.

Fig.~\ref{fig:new_pilot_msr}(b) shows a different dependence on the temporal
pilot resources. Increasing $T_s$ changes the proposed $mSR$ slightly, from approximately $0.62$ to $0.63$, while the velocity RMSE decreases from about $1.60$ to $1.51$~m/s and the position RMSE decreases
from approximately $1.0$ to $0.9$~m. The longer observation interval
provides additional temporal evolution and Doppler information, which is
more directly beneficial to velocity estimation than visual or physical localization.
Once sufficient spatial semantics have been extracted for the corresponding visual entity, additional OFDM symbols yield only marginal localization improvement. Since increasing $T_s$ also
requires more pilot observations and a longer sensing interval, the weak
$mSR$ dependence suggests that semantic sensing can operate with a
relatively short pilot observation and preserves more communication resources.

\subsection{Computational Complexity}

\begin{table}[t]
    \caption{Inference Complexity under the Dense-Scene Configuration}
    \label{tab:complexity}
    \centering
    \small
    \setlength{\tabcolsep}{5pt}
    \renewcommand{\arraystretch}{1.12}
    \begin{tabular}{lcccc}
        \toprule
        Method
        & \shortstack{Inference\\Params. (M)}
        & \shortstack{Train-only\\Params. (M)}
        & \shortstack{FLOPs\\(G)}
        & \shortstack{Latency\\(ms)}
        \\
        \midrule
        Proposed
        & 5.222 & 0.051 & 17.351 & 5.346 \\
        CNN 
        & 4.865 & 0.051 & 32.170 & 4.290 \\
        CConv 
        & 4.713 & 0.051 & 17.286 & 4.123 \\
        w/o MHSA
        & 5.024 & 0.051 & 17.344 & 5.259 \\
        w/o VRVG
        & 5.222 & 0.000 & 17.351 & 5.335 \\
        Oracle Vision
        & 4.655 & 0.000 & 16.965 & 4.177 \\
        \bottomrule
    \end{tabular}
\end{table}

Table~\ref{tab:complexity} reports the computational complexity of the
proposed framework and representative baselines. 
The reported metrics are evaluated on a server with an NVIDIA L20 GPU and an Intel Xeon Gold 6548Y+ CPUs.  
The proposed model requires $5.222$ million active parameters,
$17.351$ GFLOPs, and $5.346$ ms per scene frame.
VRVG introduces only $0.051$ million auxiliary parameters during training
and does not alter the prediction architecture or inference FLOPs. The multi-user MHSA module increases the active parameter
count by approximately $0.198$ million and adds only $0.087$ ms latency. 
Compared with the vision-only model, the proposed framework increases the
counted FLOPs by only $2.27\%$ and the measured latency by $1.169$ ms while
jointly processing all 16 user-specific radio observations.
These results indicate that the proposed radio--vision processing introduces
small computational overhead while remaining practical for real-time
inference.  
Furthermore, the output-level baselines first estimate physical parameters and then perform a separate matching stage, whereas the proposed framework directly produces user-indexed predictions and therefore avoids this additional step.

\section{Conclusions}
 
This paper developed a multimodal semantic sensing framework for OFDM-ISAC that learns user-specific radio representations oriented toward visual localization.
Gated attention compression and multi-user context learning extract task-relevant spatial structure from superimposed pilot echoes, while VRVG promotes semantic consistency with the corresponding visual entities.
The grounded radio representations act as semantic probes to retrieve and refine visual features, enabling radio--visual user matching and user-specific semantic sensing while jointly estimating position and velocity.
Experiments based on RCooper scenes and synthesized OFDM echoes demonstrated improved matching, visual localization, and physical parameter estimation relative to the evaluated baselines under interference from other users and non-user reflectors.
These findings show that semantically consistent radio representations can provide complementary information for visual inference, extending radio semantic sensing beyond physical parameter estimation toward task-oriented multimodal understanding.

\bibliographystyle{IEEEtran}
\bibliography{refs_ieee}

\vfill

\end{document}